\documentclass[10pt,journal,twoside,final,]{IEEEtran}
\usepackage{cite}
\usepackage{graphicx}
\usepackage{times}
\usepackage{latexsym}

\usepackage[mathscr]{eucal}
\usepackage{array}
\usepackage{fancyhdr}
\usepackage{amsmath,amssymb}
\usepackage{bm} 
\usepackage{subfigure}
\usepackage{multirow}

\ifCLASSINFOpdf
\else
\fi

\usepackage{xcolor}

\makeatother

\usepackage{comment}
\usepackage{amsthm}
\usepackage{amssymb}
\usepackage{graphicx}
\usepackage{tabularx}
\usepackage{color,cite}
\usepackage{booktabs}
\usepackage{url}
\usepackage{bm}
\definecolor{b}{rgb}{0,0,1}

\usepackage{algpseudocode}
\usepackage{algorithm}

\usepackage{makecell}
\usepackage{amsfonts}

\usepackage{float}
\usepackage[caption=false,font=footnotesize]{subfig}

\newtheorem{remark}{Remark}
\newtheorem{theorem}{Theorem}

\newtheorem{proposition}{Proposition}

\begin{document}

	\title{Movable Antenna-Assisted Secret Key Generation over Temporally Correlated Channels}
	
	\author{
		Zhiyu~Huang,
		Chen~Sun,~\IEEEmembership{Member,~IEEE},
		Guyue~Li,~\IEEEmembership{Member,~IEEE},  Guanghui~Wen,~\IEEEmembership{Senior Member,~IEEE}, and Xianbin Wang,~\IEEEmembership{Fellow,~IEEE}
		
		\thanks{Zhiyu Huang and Guyue Li are with the School of Cyber Science and Engineering, Southeast University, Nanjing 210096, China. Chen Sun is with the National Mobile Communications Research Laboratory, Southeast University, Nanjing 210096, China. Chen Sun and Guyue Li are also with Purple Mountain Laboratories, Nanjing 211111, China. Guyue Li is also with the Jiangsu Provincial Key Laboratory of Computer Network Technology, Nanjing 210096, China (e-mail: rain-huang-seu@seu.edu.cn; sunchen@seu.edu.cn; guyuelee@seu.edu.cn). \textit{(Corresponding author: Chen~Sun.)}}
		\thanks{Guanghui Wen is with the School of Automation, Southeast University, Nanjing 210096, China (email: ghwen@seu.edu.cn)}
		\thanks{Xianbin Wang is with the Department of Electrical and Computer Engineering, Western University, London, ON N6A 5B9, Canada (e-mail: xianbin.wang@uwo.ca).}
		\thanks{The work was supported in part by the National Natural Science Foundation of China under Grant U22A2001 and Grant 62571117. This work was supported in part by the Natural Science Foundation on Frontier Leading Technology Basic Research Project of Jiangsu under Grant BK20222001. This work was supported by the Frontier Technologies R\&D Program of Jiangsu under Grant BF2024065. This work was supported by the Postgraduate Research \& Practice Innovation Program of Jiangsu Province under Grant 26CXJH0633.}
	
}

\maketitle

\begin{abstract}
	Physical-layer key generation (PLKG) exploits reciprocal channel dynamics to generate symmetric keys, addressing the key distribution problem in wireless communications. However, strong temporal correlation in slow-fading environments limits fresh randomness and results in low key generation rates. Movable antennas (MAs) introduce spatial degrees of freedom through antenna repositioning, providing a new approach to addressing the limitation. This paper investigates MA-assisted PLKG over temporally correlated channels. A first-order Gauss-Markov process is utilized to characterize the temporal correlation of multipath gains. {We characterize the full-sequence mutual information (MI) over temporally correlated channels.
		For fixed antenna positions and beamforming, we establish a finite-horizon approximation error bound that vanishes as both effective signal-to-noise ratios (SNRs) increase. We derive a closed-form expression for the one-step conditional MI and adopt it as a tractable key-information metric. A nonconvex problem is formulated to jointly optimize antenna positions and per-slot beamforming under transmit-power and minimum-spacing constraints.
		We develop an alternating optimization algorithm, combining an exact principal-eigenvector beamforming solution with feasible SCA-based projected gradient position updates.} Simulation results demonstrate that the MA-assisted scheme outperforms the fixed-position antenna baseline. {At a high temporal correlation coefficient}, position optimization achieves approximately $51\%$ higher rates than fixed antennas. Joint position and beamforming optimization provides a further $106\%$ gain over position optimization alone, demonstrating more pronounced gains under strong temporal correlation.	
\end{abstract}

\section{Introduction}\label{sec:introduction}

With the evolution toward beyond-fifth-generation (B5G) and sixth-generation (6G) networks, future wireless systems are expected to support massive connectivity~\cite{MYNMCF21}. The growing number of devices, limited radio resources, and heterogeneous network conditions motivate lightweight security mechanisms~\cite{6GKAF2024}. Conventional upper-layer cryptographic mechanisms typically rely on computational hardness and pre-established long-term secret keys, which can become inefficient or impractical in highly dynamic wireless environments~\cite{PLS2023}. In particular, limited computational capabilities, frequent topology variations, and the absence of centralized key-management infrastructure pose fundamental challenges to secure and timely key distribution for emerging wireless devices.

Physical-layer secret key generation (PLKG) provides a promising alternative by exploiting reciprocal wireless channels to establish symmetric keys~\cite{EKB21,ZZHX22,DTAE23}. By exchanging pilots within a channel coherence interval, legitimate terminals obtain correlated observations of the same propagation channel, although receiver noise introduces discrepancies between their estimates. PLKG derives secret material from the shared propagation environment rather than pre-distributed keys, making it attractive for resource-constrained and latency-sensitive devices~\cite{LCHW17,GCXM21,JWWAT19,Maurer1993,ARCI93,HuangPLKG}. A typical PLKG procedure consists of reciprocal channel probing, quantization, information reconciliation, and privacy amplification~\cite{ZhangReview,li2018high,LXXJH21}. Quantization converts channel measurements into bit sequences, reconciliation corrects disagreements, and privacy amplification compresses the reconciled bits to limit information available to an eavesdropper, including leakage from public reconciliation. The resulting key material depends on both the quality of the reciprocal observations and the channel randomness available for extraction.

From an information-theoretic perspective, the achievable secret key rate (SKR) depends on the common randomness shared by the legitimate terminals and the information available to an eavesdropper, and is closely related to the mutual information (MI) between their channel observations~\cite{21JiTVT,LHCH26,LCZCDM25}. However, in slowly varying channels, successive probes can be strongly correlated, reducing the temporal innovation available for each new key segment and lowering the SKR~\cite{ZhangSlowFading,LiPilot25}. This limitation is particularly severe when the channel remains static or quasi-static over the coherence interval: although the legitimate terminals can still obtain high-quality observations, the amount of fresh randomness is limited.

To improve the randomness available for PLKG, existing studies have considered reconfigurable intelligent surfaces (RISs) and precoding optimization in massive multiple-input multiple-output (MIMO) systems~\cite{HLQH24,LCZCH23,22Sum-RIS,2021Sum,LMZW25,mimo25tw}. RIS-based designs can require deploying and controlling additional reflecting elements, while precoding shapes observations using a fixed array geometry. Depending on the architecture, additional hardware, control, or energy overhead accompanies these approaches. This motivates a complementary degree of freedom that directly adjusts the spatial channel response at the legitimate transmitter without relying on additional reflecting infrastructure.

Movable antenna (MA) technology provides this degree of freedom~\cite{MZSZ26}. Unlike fixed-position antenna (FPA) arrays, MA arrays allow element locations to be adjusted within a prescribed region, making antenna position an additional design variable~\cite{ZhuMA23,MATutorial25,6DQDLSZ26}. Under the field-response model, antenna coordinates alter the phases of resolvable multipath components and thereby reshape the spatial channel response and correlation structure~\cite{ModelingMA23,xlgyxt26,CZGWNZWZ26,BZCLYJS26,MaMIMO24,YZZ26}. Even when propagation angles and path-gain statistics remain unchanged over the movable region, different layouts produce different combinations of the multipath components at the antenna elements. Position selection consequently determines the spatial observations available to the array, while beamforming controls how these observations are combined. Their joint design allows the effective channel to be adjusted through both antenna geometry and signal processing.

MA reconfiguration has been studied in secure transmission~\cite{ChengSecure24,HuSPL24,TangMIMO25,ChengNoCSI25,HuTMC24,FengPLS25}, covert communication~\cite{LiuCovert25,MaoCovert25}, and integrated sensing and communication~\cite{LeISAC25,BSPL26}. For PLKG, position selection offers a means to improve the observation of available channel fluctuations, even if the selected layout remains fixed during probing. The relevant information then depends on both the current observation quality and its dependence on earlier probes. In~\cite{Matsumine25}, random antenna repositioning introduces additional randomness in quasi-static environments with fixed propagation parameters. {However, the setting does not address temporally evolving path gains or the interaction with joint multi-antenna position and beamforming optimization. A configuration that maximizes the MI in one slot does not necessarily maximize the additional shared information available in the next slot. Temporal correlation creates dependence between consecutive observation pairs. Accounting for the preceding observations is necessary to evaluate the additional shared information provided by the next probe.}
{Addressing the problem involves two coupled technical challenges. The first challenge is to derive a tractable conditional MI expression that accounts for temporal dependence and to quantify its approximation error relative to the full-sequence joint MI. Second, antenna positions and consecutive-slot beamforming jointly determine both the current channel power and the component predictable from earlier observations. This coupling produces a nonconvex objective, while minimum-spacing constraints further restrict feasible position updates.}

In this paper, we investigate MA-assisted PLKG over temporally correlated channels. {We jointly design the antenna layout and per-slot beamforming to improve the utilization of available channel variation under a transmit-power constraint, with the selected layout held fixed during probing.} The main contributions of this paper are summarized as follows:

\begin{itemize}
	\item {We develop an MA-assisted PLKG model with first-order Gauss--Markov multipath gains and derive the one-step conditional MI in closed form. {The expression quantifies the shared information in the current observation pair conditioned on the preceding pair, capturing the coupled effects of temporal correlation, antenna positions, and consecutive-slot beamforming. For fixed antenna positions and beamforming and a fixed temporal correlation coefficient, we bound the finite-horizon approximation error defined in Theorem~\ref{thm:finite_horizon_approx} and prove that the error vanishes as both effective SNRs increase.}}
	
	\item We jointly optimize antenna positions and per-slot beamforming under transmit-power, movable-region, and minimum-spacing constraints. {By expressing the conditional rate through the residual channel covariance, we obtain an exact principal-eigenvector solution to the beamforming subproblem that accounts for both current channel power and dependence on the preceding observation pair. We develop an alternating optimization (AO) algorithm combining this solution with projected gradient ascent for antenna positioning, derive the analytical position gradient, and use successive convex approximation (SCA) to preserve antenna separation. Backtracking ensures that accepted position updates do not decrease the objective.}
	
	\item {Through simulations, we analyze the effects of temporal correlation, antenna number, and transmit power. The comparisons demonstrate the gains of antenna-position optimization over the FPA baseline and the further gains of joint position and beamforming design under strong temporal correlation. We also compare the one-step conditional MI with the finite-horizon joint MI and block-wise approximation to examine its accuracy under the considered observation conditions.}
\end{itemize}


\emph{Notations:}
In this paper, matrices and vectors are denoted by boldface upper-case and boldface lower-case, respectively.
$\mathbb{C}^{A\times B}$ and $\mathbb{R}^{A\times B}$ denote the spaces of complex and real matrices of size $A\times B$, respectively.
$(\cdot)^*$, $(\cdot)^{\top}$, and $(\cdot)^{\mathsf{H}}$ denote the conjugate, transpose, and conjugate transpose, respectively.
$\operatorname{diag}(\boldsymbol{x})$ is a matrix whose main diagonal elements are the entries of $\boldsymbol{x}$.
$\operatorname{det}(\cdot)$ is the matrix determinant.
$\|\boldsymbol{x}\|_2$ denotes the $\ell_2$ norm of vector $\boldsymbol{x}$.
$\|\cdot\|_{\mathrm F}$ denotes the Frobenius norm.
$\mathbb{E}\{\cdot\}$ represents statistical expectation.
$\boldsymbol{b}\sim \mathcal{CN}\left(\boldsymbol{0}, \boldsymbol{\Sigma}\right)$ denotes that $\boldsymbol{b}$ is a circularly symmetric complex Gaussian (CSCG) vector with zero mean and covariance matrix $\boldsymbol{\Sigma}$.
$\mathcal{I}\left(X; Y\right)$ is the mutual information between $X$ and $Y$.
$\mathrm{Re}\{\cdot\}$ represents the real part of a complex number.
$\mathcal{O}(\cdot)$ denotes the computational complexity.

\section{System Model}\label{sec:system_model}
\begin{figure}[htb]
	\centering
	\includegraphics[width=\columnwidth]{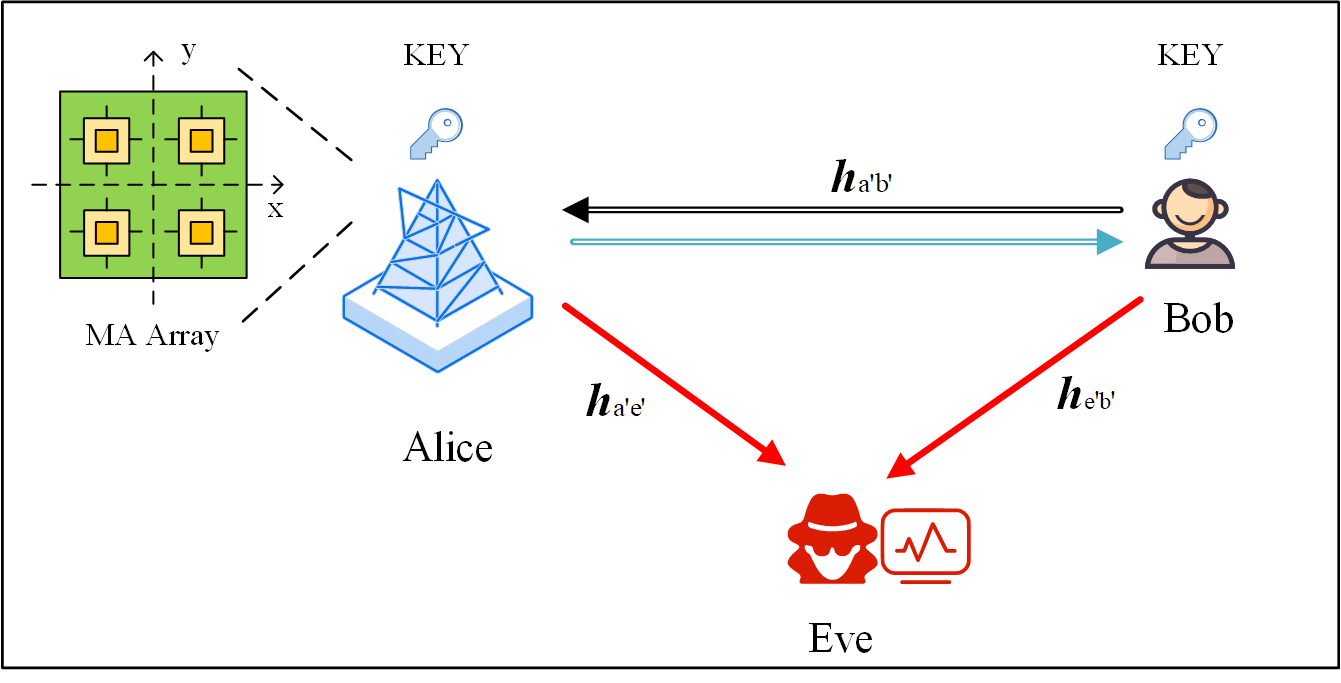}
	\caption{MA-assisted secret key generation over a point-to-point reciprocal slow-fading channel.}
	\label{system model}
\end{figure}

{Fig.~\ref{system model} depicts a point-to-point reciprocal wireless link between Alice, equipped with $N$ movable antennas, and single-antenna Bob. To encrypt confidential communications, Alice and Bob perform channel probing and establish symmetric secret keys from reciprocal channel observations. The antenna layout is selected before the analyzed probing interval and remains fixed throughout the interval.\footnote{A time-division duplexing (TDD) protocol is adopted, with the two probing phases within each slot completed over the same channel realization~\cite{LCZCH23,HLQH24}.} }
\subsection{Channel Model}

{We consider a narrowband multipath channel with $N_p$ resolvable plane-wave components under TDD channel reciprocity.}
{Alice employs $N$ movable antennas (MAs), whose positions can be adjusted within a prescribed region. Let $\mathbf U=[\mathbf u_1,\ldots,\mathbf u_N]\in\mathbb R^{2\times N}$ collect the MA positions, where $\mathbf u_n=[x_n,y_n]^\top$. Each antenna moves within the square region $\mathcal C=[0,A]\times[0,A]$. To limit inter-antenna coupling, the positions satisfy}
\begin{align}
	\|\mathbf u_i-\mathbf u_j\|_2
	&\geq d_{\min},\quad i\neq j,
	\label{eq:spacing}\\
	x_n,y_n
	&\in[0,A],\quad n=1,\ldots,N,
	\label{eq:region}
\end{align}
{where $d_{\min}$ is the minimum antenna spacing, typically set to half the carrier wavelength $\lambda$. All $N$ antennas are active.}\footnote{{Unlike FPA arrays with fixed element positions, the MA layout can be re-optimized when the temporal correlation coefficient changes.}}
{
	{Let $\theta_p$ and $\phi_p$ denote the elevation and azimuth angles of departure of path $p$}~\cite{PlaneWave26,ModelingMA23}.
	For path $p$, the field-response model gives the MA-position-dependent steering vector
	\begin{align}
		\boldsymbol a(\theta_p,\phi_p,\mathbf U)
		=\frac{1}{\sqrt N}\left[
		e^{j\frac{2\pi}{\lambda}\mathbf u_1^\top\boldsymbol\rho_p},\ldots,
		e^{j\frac{2\pi}{\lambda}\mathbf u_N^\top\boldsymbol\rho_p}
		\right]^\top, \label{eq:steering_2d}
	\end{align}
	where $\boldsymbol\rho_p=[\sin\theta_p\cos\phi_p,\cos\theta_p]^\top$ contains the in-plane direction cosines. Let $\beta_{t,p}$ denote the complex gain of path $p$ in slot $t$. The reciprocal channel vector is the sum of the path responses,
	\begin{align}
		\boldsymbol h_t(\mathbf U)
		=\sum_{p=1}^{N_p}\beta_{t,p}\boldsymbol a(\theta_p,\phi_p,\mathbf U), \label{eq:path_sum}
	\end{align}
	{where $\boldsymbol h_t^{\mathrm{UL}}\triangleq\boldsymbol h_t(\mathbf U)$ and $\boldsymbol h_t^{\mathrm{DL}}\triangleq\boldsymbol h_t(\mathbf U)^{\mathsf H}$.}}

{Collecting the path responses and path gains into $\mathbf A(\mathbf U)=[\boldsymbol a(\theta_1,\phi_1,\mathbf U),\ldots,\boldsymbol a(\theta_{N_p},\phi_{N_p},\mathbf U)]$ and $\boldsymbol b_t=[\beta_{t,1},\ldots,\beta_{t,N_p}]^\top$, respectively, gives
	\begin{align}
		\boldsymbol h_t(\mathbf U)=\mathbf A(\mathbf U)\boldsymbol b_t\in\mathbb C^{N\times1}. \label{eq:channel_matrix}
	\end{align}
}
{In slow-fading environments, successive channel observations can be temporally correlated. Accordingly, the path-gain vector follows a first-order Gauss--Markov model~\cite{2021Sum,Zhangjun_cor}:}

\begin{align}
	\boldsymbol b_t=\alpha\boldsymbol b_{t-1}
	+\sqrt{1-\alpha^2}\,\boldsymbol\epsilon_t,\quad t\geq1, \label{eq:markov_channel}
\end{align}
{where $0\leq\alpha<1$ denotes the one-step temporal correlation coefficient. 
The initial path-gain $\boldsymbol b_0$ and the innovation $\boldsymbol\epsilon_t$ are mutually independent, each following \(\mathcal{CN}(\boldsymbol 0,\sigma_b^2\mathbf I_{N_p})\).}
{
	For a fixed antenna layout $\mathbf U$, the same-slot and cross-slot channel statistics are given by
	\begin{align}
		\mathbb E\!\left\{
		\boldsymbol h_{t_1}(\mathbf U)
		\boldsymbol h_{t_2}(\mathbf U)^{\mathsf H}
		\right\}
		&=
		\alpha^{|t_1-t_2|}
		\bar{\mathbf R}(\mathbf U),
		\label{eq:cross_corr}\\
		\bar{\mathbf R}(\mathbf U)
		&\triangleq
		\sigma_b^2\mathbf A(\mathbf U)
		\mathbf A(\mathbf U)^{\mathsf H}.
		\label{eq:bar_R}
	\end{align}
	where $t_1,t_2\in\{1,\ldots,T\}$. The case $t_1=t_2$ corresponds to the spatial covariance, while adjacent slots have cross-covariance $\alpha\bar{\mathbf R}(\mathbf U)$. Accordingly, the channel innovation $\sqrt{1-\alpha^2}\mathbf A(\mathbf U)\boldsymbol\epsilon_t$ has covariance $(1-\alpha^2)\bar{\mathbf R}(\mathbf U)$.
	The $(i,j)$-th entry of the spatial covariance can be expressed as
	\begin{align}
		[\bar{\mathbf R}(\mathbf U)]_{i,j}
		=\frac{\sigma_b^2}{N}\sum_{p=1}^{N_p}
		e^{j\frac{2\pi}{\lambda}(\mathbf u_i-\mathbf u_j)^\top\boldsymbol\rho_p}. \label{eq:spatial_cov_entries}
	\end{align}
}
\begin{remark}
	The diagonal entries in \eqref{eq:spatial_cov_entries} equal $N_p\sigma_b^2/N$, whereas the off-diagonal entries depend on the relative antenna coordinates. Under the adopted unit-norm steering-vector normalization, $\operatorname{tr}(\bar{\mathbf R}(\mathbf U))=N_p\sigma_b^2$ is independent of the antenna layout. Position optimization reshapes the spatial covariance structure without changing the total average channel energy.
\end{remark}

\subsection{Signal Model}

{During the secret key generation}, Alice and Bob alternately transmit unit-modulus pilot symbols $s_{\mathrm d}$ and $s_{\mathrm u}$ in each slot. During the Alice-to-Bob phase, Alice precodes $s_{\mathrm d}$ with $\boldsymbol w_t\in\mathbb C^{N\times1}$, where $\|\boldsymbol w_t\|_2^2\leq P_{\mathrm A}$ and $P_{\mathrm A}$ denotes the transmit-power budget. 
Bob receives 
\begin{equation}
	y_{\mathrm b,t}^{\mathrm d}=\boldsymbol h_t^{\mathsf H}(\mathbf U)\boldsymbol w_t s_{\mathrm d}+z_{\mathrm b,t}^{\mathrm d},
\end{equation}
{where $z_{\mathrm b,t}^{\mathrm d}\sim\mathcal{CN}(0,\sigma_z^2)$. The LS estimate is used to align Bob's observation with Alice's reciprocal observation~\cite{21JiTVT}.}

\begin{align}
	\hat h_{\mathrm b,t}
	&\triangleq(s_{\mathrm d}^*y_{\mathrm b,t}^{\mathrm d})^*
	=\boldsymbol w_t^{\mathsf H}\mathbf A(\mathbf U)\boldsymbol b_t
	+\tilde z_{\mathrm b,t}^{\mathrm d}, \label{eq:hb_obs}
\end{align}
where $\tilde z_{\mathrm b,t}^{\mathrm d}=(s_{\mathrm d}^*z_{\mathrm b,t}^{\mathrm d})^*\sim\mathcal{CN}(0,\sigma_z^2)$.

During the Bob-to-Alice phase, Bob transmits $s_{\mathrm u}$ at power $P_{\mathrm B}$, yielding
\begin{align}
	\boldsymbol y_{\mathrm a,t}^{\mathrm u}
	=\sqrt{P_{\mathrm B}}\,\boldsymbol h_t(\mathbf U)s_{\mathrm u}
	+\boldsymbol z_{\mathrm a,t}^{\mathrm u},
\end{align}
where $\boldsymbol z_{\mathrm a,t}^{\mathrm u}\sim\mathcal{CN}(\boldsymbol0,\sigma_z^2\mathbf I_N)$. Alice combines the LS estimate using the same $\boldsymbol w_t$:
\begin{align}
	\hat h_{\mathrm a,t}
	&\triangleq\boldsymbol w_t^{\mathsf H}(s_{\mathrm u}^*\boldsymbol y_{\mathrm a,t}^{\mathrm u})\notag\\
	&=\sqrt{P_{\mathrm B}}\,\boldsymbol w_t^{\mathsf H}\mathbf A(\mathbf U)\boldsymbol b_t
	+z_{\mathrm a,t}^{\mathrm u}, \label{eq:ha_obs}
\end{align}
where $z_{\mathrm a,t}^{\mathrm u}=s_{\mathrm u}^*\boldsymbol w_t^{\mathsf H}\boldsymbol z_{\mathrm a,t}^{\mathrm u}$ has variance {$\|\boldsymbol w_t\|_2^2\sigma_z^2$}. The receiver noises are independent across terminals and probing slots and are independent of the channel process. The resulting observation pairs are subsequently used for quantization, information reconciliation, and privacy amplification~\cite{2021Sum,li2018high}.

{
A passive Eve equipped with a single FPA overhears the pilot transmissions in both probing phases. Let \(\boldsymbol h_{\mathrm{ae},t}(\mathbf U)\in\mathbb C^{N\times1}\) and \(h_{\mathrm{be},t}\in\mathbb C\) denote the Alice–Eve and Bob–Eve channels, respectively. Based on the received pilot signals, Eve obtains noisy observations of the corresponding effective channels. After pilot removal in both phases and conjugation of the Alice-to-Bob observation, these observations are expressed as
\begin{align}
	\hat h_{\mathrm e,t}^{\mathrm d}
	&=\boldsymbol w_t^{\mathsf H}\boldsymbol h_{\mathrm{ae},t}(\mathbf U)
	+\tilde z_{\mathrm e,t}^{\mathrm d},\\
	\hat h_{\mathrm e,t}^{\mathrm u}
	&=\sqrt{P_{\mathrm B}}\,h_{\mathrm{be},t}
	+z_{\mathrm e,t}^{\mathrm u},
\end{align}
where \(\tilde z_{\mathrm e,t}^{\mathrm d}\) and \(z_{\mathrm e,t}^{\mathrm u}\) denote independent Gaussian receiver noises at Eve. The observation pair is denoted by \(E_t=(\hat h_{\mathrm e,t}^{\mathrm d},\hat h_{\mathrm e,t}^{\mathrm u})\).
}
\section{Secret Key Rate Analysis}\label{sec:skr_model}
{In this section, we analyze the key-information metrics over temporally correlated channels using mutual information (MI). Under the adopted spatial-decorrelation assumption for Eve, information leakage is considered negligible. We adopt the full-sequence joint MI as a reference and introduce block-wise and one-step conditional MI approximations to simplify the analysis, followed by an approximation error bound and a closed-form expression for the latter.}

\subsection{Full-Sequence Benchmark and Block-Wise Approximation}\label{sec:mi_comparison}

{To account for temporal correlation across probing slots, we analyze the secret key rate using the joint mutual information (MI) of the complete observation sequences. Specifically, let \(A_t=\hat h_{\mathrm a,t}\), \(B_t=\hat h_{\mathrm b,t}\), \(A^T=(A_1,\ldots,A_T)\), \(B^T=(B_1,\ldots,B_T)\), and \(E^T=(E_1,\ldots,E_T)\). For independent repetitions of the length-\(T\) observation vectors with authenticated public discussion, an achievable rate normalized by \(T\) is given by~\cite{Maurer1993,ARCI93}}
\begin{align}
	R_T\triangleq\frac{1}{T}
	\big[\mathcal I(A^T;B^T)-\mathcal I(A^T;E^T)\big]^+,
	\label{eq:sequence_skr_bounds}
\end{align}
where $[x]^+=\max\{x,0\}$ {and} the second MI term quantifies the information leakage to Eve.

{Following the spatial-decorrelation assumption for Eve~\cite{Matsumine25}, we assume that Eve is located more than half a wavelength away from Bob and MA position at Alice. The eavesdropping channel processes are modeled as independent of the legitimate channel process, while receiver noises are mutually independent and independent of the channel processes. Consequently, $E^T$ is independent of the legitimate observation sequences $(A^T,B^T)$ under the adopted model.}

{Under the stated process-level independence assumption, $\mathcal I(A^T;E^T)\approx0$, and~\eqref{eq:sequence_skr_bounds} reduces to the finite-horizon joint MI}
\begin{align}
	R_T^{\mathrm{joint}}\triangleq\frac{1}{T}\mathcal I(A^T;B^T).
	\label{eq:global_mi_rate}
\end{align}

The finite-horizon joint MI $R_T^{\mathrm{joint}}$ retains all temporal dependence in the length-$T$ observation sequences. Its evaluation involves an observation dimension that grows with $T$, making repeated evaluation within antenna-position and beamforming optimization increasingly costly.
A conventional way to reduce the dimension is to divide the observations into shorter blocks. For $T=KL$, let $\mathcal T_k=\{(k-1)L+1,\ldots,kL\}$ for $k=1,\ldots,K$, and let $A_{\mathcal T_k}$ and $B_{\mathcal T_k}$ collect the corresponding observations. The block-wise approximation is defined as
\begin{align}
	R_{T,L}^{\mathrm{block}}\triangleq\frac{1}{T}\sum_{k=1}^{K}
	\mathcal I(A_{\mathcal T_k};B_{\mathcal T_k}). \label{eq:block_mi_rate}
\end{align}

It preserves correlation within each block but neglects dependence across block boundaries. Increasing $L$ retains more temporal dependence at a higher evaluation cost, whereas short blocks discard information dependence that remains significant in slowly varying channels.

\begin{figure}[htb]
	\centering
	\includegraphics[width=\columnwidth]{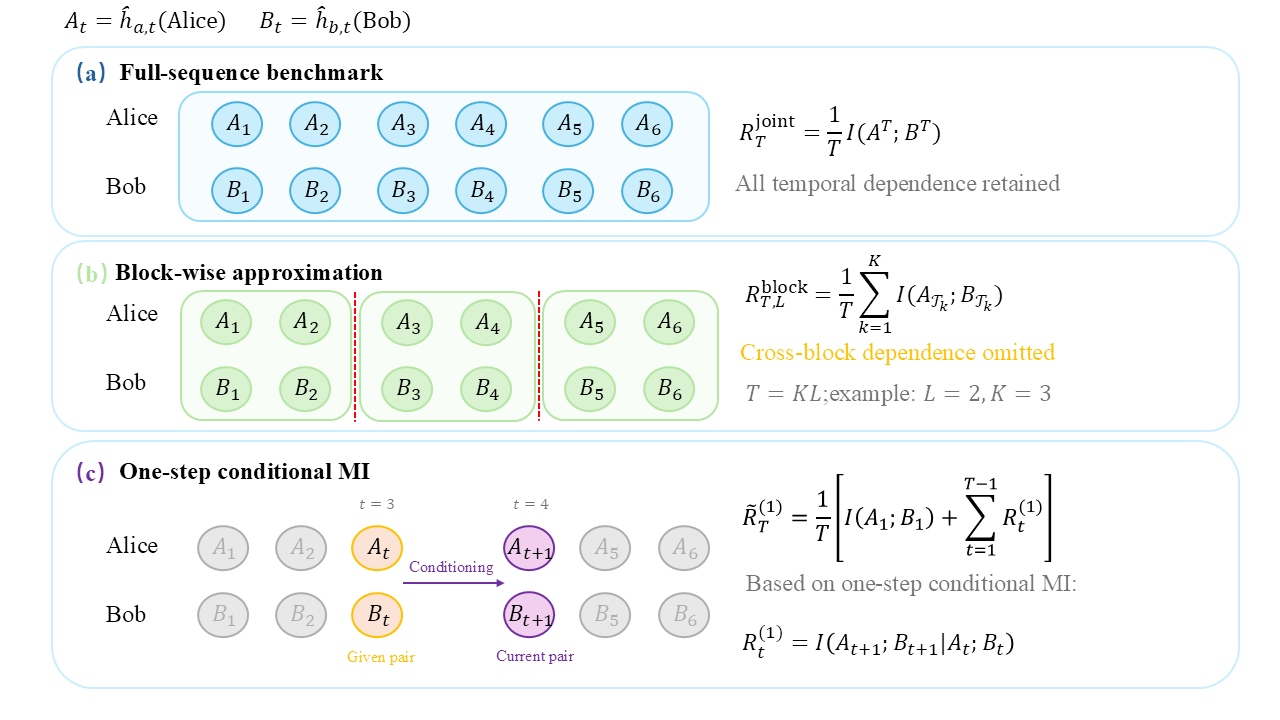}
	\caption{{Observation dependence retained by (a) the finite-horizon joint MI benchmark, (b) the block-wise approximation, and (c) the one-step conditional MI.}}
	\label{model_comparison}
\end{figure}

{The joint MI benchmark, block-wise approximation, and one-step formulation are illustrated in Fig.~\ref{model_comparison}.}

\subsection{One-Step Rate and High-SNR Consistency}

{
To reduce the complexity of evaluating the full-sequence MI, we consider a fixed-window conditional MI based on consecutive probing slots. Under the first-order Gauss–Markov model, the current channel state depends on its history only through the preceding state. 
However, due to receiver noise, the channel observations do not generally inherit this Markov property, and their temporal dependence may extend beyond the preceding slot. Specifically, the chain rule gives $\mathcal I(A^T;B^T)=\sum_{t=1}^{T}\mathcal I(A_t;B^T\mid A^{t-1})$, where the observation histories remain involved. We therefore condition each new observation pair on the preceding pair to obtain a tractable per-probe metric.
}

We use the adjacent-pair conditional information $R_t^{(1)}$ as the per-slot metric, which is defined as
\begin{align}
	R_t^{(1)}\triangleq\mathcal I(A_{t+1};B_{t+1}\mid A_t,B_t),
	\quad t=1,\ldots,T-1, \label{eq:skr_approx}
\end{align}

The following theorem bounds the aggregate terms omitted by this history truncation.

\begin{theorem}[{Finite-Horizon Approximation and High-SNR Consistency}]\label{thm:finite_horizon_approx}
	Suppose that $\mathbf U$ and a nonzero $\boldsymbol w$ are fixed over $T\geq2$ slots, with $0\leq\alpha<1$ and $S=\boldsymbol w^{\mathsf H}\bar{\mathbf R}(\mathbf U)\boldsymbol w>0$. Let $r_{\mathrm a}=\|\boldsymbol w\|_2^2\sigma_z^2/P_{\mathrm B}$, $r_{\mathrm b}=\sigma_z^2$, and $\gamma_{\mathrm x}=S/r_{\mathrm x}$ for $\mathrm x\in\{\mathrm a,\mathrm b\}$. Define
	\begin{align}
		\bar R_T^{(1)}&\triangleq\frac{1}{T-1}\sum_{t=1}^{T-1}\mathcal I(A_{t+1};B_{t+1}\mid A_t,B_t),\notag\\
		\mathcal E_T&\triangleq\left|R_T^{\mathrm{joint}}-\frac{\mathcal I(A_1;B_1)}{T}-\frac{T-1}{T}\bar R_T^{(1)}\right|. \label{eq:finite_horizon_error}
	\end{align}
	Then
	\begin{align}
		g_\alpha(\gamma)&\triangleq\log_2\!\left(1+\frac{\alpha^2}{1+(1-\alpha^2)\gamma}\right), \label{eq:g_alpha_def}\\
		\mathcal E_T&\leq\frac{T-1}{T}\left[g_\alpha(\gamma_{\mathrm a})+g_\alpha(\gamma_{\mathrm b})+g_\alpha(\gamma_{\mathrm a}+\gamma_{\mathrm b})\right], \label{eq:high_snr_error_bound}\\
		\lim_{\gamma_{\mathrm a},\gamma_{\mathrm b}\to\infty}\mathcal E_T&=0,\quad\text{for fixed }\alpha<1. \label{eq:high_snr_limit}
	\end{align}
\end{theorem}
\begin{proof}
	{The proof is given in Appendix~\ref{app:finite_horizon_approx}.}
\end{proof}

\begingroup
The following result extends the information-increment analysis to an adjacent pair with distinct beamformers.
\begin{proposition}[Two-Slot Extension with Distinct Beamformers]\label{prop:two_slot_extension}
	Under the Gaussian observation model with $P_{\mathrm B},\sigma_z^2>0$ and fixed $0\leq\alpha<1$, let $\mathbf U$ be fixed over slots $t$ and $t+1$. For deterministic beamformers with $S_i=\boldsymbol w_i^{\mathsf H}\bar{\mathbf R}(\mathbf U)\boldsymbol w_i>0$, $i\in\{t,t+1\}$, define the reference-slot SNRs $\gamma_{\mathrm a,t}=P_{\mathrm B}S_t/(\|\boldsymbol w_t\|_2^2\sigma_z^2)$ and $\gamma_{\mathrm b,t}=S_t/\sigma_z^2$, and the two-slot joint MI increment
	\begin{align}
		\Delta_t^{(2)}
		&\triangleq\mathcal I(A_t,A_{t+1};B_t,B_{t+1})\notag\\
		&\quad-\mathcal I(A_t;B_t). \label{eq:two_slot_increment}
	\end{align}
	Then, for $R_t^{(1)}$ in~\eqref{eq:skr_approx},
	\begin{align}
		0&\leq\Delta_t^{(2)}-R_t^{(1)}\notag\\
		&\leq\sum_{\mathrm x\in\{\mathrm a,\mathrm b\}}
		\log_2\!\left(1+\frac{\alpha^2}
		{(1-\alpha^2)(1+\gamma_{\mathrm x,t})}\right). \label{eq:two_slot_error_bound}
	\end{align}
	Thus, $\Delta_t^{(2)}-R_t^{(1)}\to0$ as both reference-slot SNRs tend to infinity, even when $\boldsymbol w_{t+1}\neq\boldsymbol w_t$.
\end{proposition}
\begin{proof}
	The chain rule gives $\Delta_t^{(2)}-R_t^{(1)}=\mathcal I(A_{t+1};B_t\mid A_t)+\mathcal I(B_{t+1};A_t\mid B_t)\geq0$. For the upper bound, rescale Alice's observations by $1/\sqrt{P_{\mathrm B}}$, retaining the same symbols. Write $G_i=\boldsymbol w_i^{\mathsf H}\boldsymbol h_i(\mathbf U)$ and $r_{\mathrm a,i}=\|\boldsymbol w_i\|_2^2\sigma_z^2/P_{\mathrm B}$. By~\eqref{eq:cross_corr} and covariance Cauchy--Schwarz,
	$\zeta=|\mathbb E\{G_{t+1}G_t^*\}|^2/(S_tS_{t+1})\leq\alpha^2$.
	Set $\mathcal{V}_{\mathrm a}=(1-\zeta)S_{t+1}+r_{\mathrm a,t+1}$. Gaussian conditioning gives
	$\operatorname{Var}(A_{t+1}\mid A_t)=\mathcal{V}_{\mathrm a}+\zeta S_{t+1}/(1+\gamma_{\mathrm a,t})$ and
	$\operatorname{Var}(A_{t+1}\mid A_t,B_t)\geq \mathcal{V}_{\mathrm a}$.
	Hence,
	\begin{align}
		\mathcal I(A_{t+1};B_t\mid A_t)
		&\leq\log_2\!\left(1+\frac{\zeta S_{t+1}}
		{(1+\gamma_{\mathrm a,t})\mathcal V_{\mathrm a}}\right)\notag\\
		&\leq\log_2\!\left(1+\frac{\alpha^2}
		{(1-\alpha^2)(1+\gamma_{\mathrm a,t})}\right).\notag
	\end{align}
	The symmetric bound for $\mathcal I(B_{t+1};A_t\mid B_t)$ completes the proof.
\end{proof}
\endgroup

\begin{remark}
	{Theorem~\ref{thm:finite_horizon_approx} establishes finite-horizon high-SNR consistency for fixed antenna positions and beamforming. For two adjacent slots with possibly different beamformers, Proposition~\ref{prop:two_slot_extension} establishes $R_t^{(1)}\leq\Delta_t^{(2)}$ and shows that the gap vanishes as both reference-slot effective SNRs tend to infinity for fixed $\alpha<1$. Thus, the conditional MI is an asymptotically tight lower bound on the additional joint information contributed by the second probing slot. This result motivates the subsequent maximization of that lower bound over $\mathbf U$ and $\boldsymbol w_{t+1}$ for a prescribed $\boldsymbol w_t$. The design is performed before either slot is probed, with a common layout across the pair. The guarantee applies to the analyzed adjacent pair.}
\end{remark}

{
We next derive a closed-form expression for the one-step conditional MI. For two consecutive probing slots, we approximate the per-slot SKR by \(R_{\mathrm{SK}}(\boldsymbol w_{t+1},\mathbf U)\approx R_t^{(1)}\), which depends on the current-slot beamforming and antenna positions.
}
\begin{align}
	R_{\mathrm{SK}}(\boldsymbol w_{t+1},\mathbf U)
	=\log_2\frac{\det(\mathbf R_{\mathrm{aa'b'}})\det(\mathbf R_{\mathrm{ba'b'}})}
	{\det(\mathbf R_{\mathrm{aba'b'}})\det(\mathbf R_{\mathrm{a'b'}})}, \label{eq:det_formula}
\end{align}
where each covariance matrix contains the observations indicated by the subscript. The explicit structures are given in Appendix~\ref{sec:cov_matrices}.
{
Let $p=\boldsymbol w_t^{\mathsf H}\bar{\mathbf R}(\mathbf U)\boldsymbol w_t,		q=\boldsymbol w_{t+1}^{\mathsf H}\bar{\mathbf R}(\mathbf U)\boldsymbol w_{t+1},
c=\alpha^2|\boldsymbol w_t^{\mathsf H}\bar{\mathbf R}(\mathbf U)\boldsymbol w_{t+1}|^2$. The determinant of the preceding-pair covariance matrix is
	\begin{align}
		D&\triangleq\det(\mathbf R_{\mathrm{a'b'}})\notag\\
		&=\sigma_z^2\big[(P_{\mathrm B}+\|\boldsymbol w_t\|_2^2)p
		+\|\boldsymbol w_t\|_2^2\sigma_z^2\big]. \label{eq:D_def}
	\end{align}
	
	Let $K_{\mathrm b}=\sigma_z^2(P_{\mathrm B}+\|\boldsymbol w_t\|_2^2)$, $K_{\mathrm a}=P_{\mathrm B}K_{\mathrm b}$, and $K_{\mathrm{ab}}=\sqrt{P_{\mathrm B}}K_{\mathrm b}$. To simplify~\eqref{eq:det_formula}, define $
		{V_{\mathrm a}}=P_{\mathrm B}q+\|\boldsymbol w_{t+1}\|_2^2\sigma_z^2, 
		{V_{\mathrm b}}=q+\sigma_z^2, 
		{C_{\mathrm{ab}}}=\sqrt{P_{\mathrm B}}q$.
	The Schur-complement expressions for the determinants in~\eqref{eq:det_formula} are
	\begin{align}
		\det(\mathbf R_{\mathrm{aa'b'}})&=D {V_{\mathrm a}}-K_{\mathrm a}c,\label{eq:det_Raa'b'}\\
		\det(\mathbf R_{\mathrm{ba'b'}})&=D {V_{\mathrm b}}-K_{\mathrm b}c,\label{eq:det_Rba'b'}\\
		\det(\mathbf R_{\mathrm{aba'b'}})&=D\Big[
		\left({V_{\mathrm a}}-\frac{K_{\mathrm a}c}{D}\right)
		\left({V_{\mathrm b}}-\frac{K_{\mathrm b}c}{D}\right)\notag\\
		&\qquad-\left({C_{\mathrm{ab}}}-\frac{K_{\mathrm{ab}}c}{D}\right)^2\Big].\label{eq:det_Raba'b'}
	\end{align}
Conditioned on the preceding observation pair $(A_t,B_t)$,
the covariance matrix of $(A_{t+1},B_{t+1})$ is given by
\begin{align}
	\boldsymbol{\Sigma}_{t+1|t}
	=
	\begin{bmatrix}
		D_1 & \mathcal N\\
		\mathcal N^* & D_2
	\end{bmatrix},
\end{align}
where
	\begin{align}
		\mathcal N(\boldsymbol w_{t+1},\mathbf U)
		&={C_{\mathrm{ab}}}-\frac{K_{\mathrm{ab}}c}{D}, \label{eq:N_def}\\
		D_1(\boldsymbol w_{t+1},\mathbf U)
		&={V_{\mathrm a}}-\frac{K_{\mathrm a}c}{D}, \label{eq:D1_def}\\
		D_2(\boldsymbol w_{t+1},\mathbf U)
		&={V_{\mathrm b}}-\frac{K_{\mathrm b}c}{D}. \label{eq:D2_def}
	\end{align}
	Inserting these entries into the Gaussian MI formula gives\footnote{For $\alpha=0$, \eqref{eq:skr_compact} reduces to the MI
		between the current observations, as consecutive probing pairs
		are independent.}
	\begin{align}
		R_{\mathrm{SK}}(\boldsymbol w_{t+1},\mathbf U)
		&=\log_2\!\left(\frac{1}{1-X(\boldsymbol w_{t+1},\mathbf U)}\right), \label{eq:skr_compact}\\
		X(\boldsymbol w_{t+1},\mathbf U)
		&=\frac{\mathcal N^2(\boldsymbol w_{t+1},\mathbf U)}
		{D_1(\boldsymbol w_{t+1},\mathbf U)D_2(\boldsymbol w_{t+1},\mathbf U)}. \label{eq:X_def}
	\end{align}
	{For nonzero consecutive-slot beamforming and positive receiver-noise variance, the conditional observation covariance is positive definite and $0\leq X<1$.}
}

\section{{Joint Beamforming and Antenna Position Optimization for Key-Information Maximization}}\label{sec:optimization}

{
In this section, we jointly optimize the antenna positions
and per-slot beamforming to maximize the one-step conditional
key-information rate. We first reformulate the optimization
problem and then develop an alternating optimization algorithm.
As shown in Eq.~\eqref{eq:skr_compact}, the one-step rate is
jointly determined by the array covariance shaped by $\mathbf U$
and the current beamforming $\boldsymbol w_{t+1}$, through
both current channel power and cross-slot dependence.
Since $R_{\mathrm{SK}}$ is strictly increasing in
$X\in(0,1)$, maximizing~\eqref{eq:skr_compact}
is equivalent to the following fractional problem:
}

\begin{subequations}\label{problem:P}
	\begin{align}
		\max_{\boldsymbol w_{t+1} \in \mathbb{C}^{N \times 1}, \; \mathbf{U} \in \mathbb{R}^{2 \times N}} \quad & f(\boldsymbol w_{t+1}, \mathbf{U}) \triangleq X(\boldsymbol w_{t+1}, \mathbf{U}) \label{problem:P_obj} \\
		\text{s.t.} \quad & \|\boldsymbol w_{t+1}\|_2^2 \leq P_{\mathrm{A}}, \label{problem:P_c1} \\
		& \|\mathbf{u}_i - \mathbf{u}_j\|_2 \geq d_{\min}, \quad \forall i \neq j, \label{problem:P_c2} \\
		& x_n \in [0, A], \; y_n \in [0, A], \quad \forall n, \label{problem:P_c3}
	\end{align}
\end{subequations}

{{The objective in~\eqref{problem:P} is non-concave in $\boldsymbol w_{t+1}$ and $\mathbf U$, which are coupled through $\bar{\mathbf R}(\mathbf U)$, and the minimum-spacing constraints are non-convex. To solve the problem, an alternating optimization (AO) method is proposed to update the beamforming and the antenna positions.}}

\subsection{Beamforming Optimization}
\begingroup
Given antenna positions $\mathbf U$ and a nonzero preceding-slot beamforming $\boldsymbol w_t$, the beamforming subproblem is
\begin{align}
	\max_{\boldsymbol w_{t+1}}\quad&
	X(\boldsymbol w_{t+1},\mathbf U)\notag\\
	\text{s.t.}\quad&\|\boldsymbol w_{t+1}\|_2^2\leq P_{\mathrm A}.
	\label{eq:sub1}
\end{align}

{To separate transmit power from spatial weighting, the nonzero beamforming vector can be decomposed as $\boldsymbol w_{t+1}=\sqrt{\rho}\boldsymbol v$, where $\rho=\|\boldsymbol w_{t+1}\|_2^2$ denotes the transmit power and $\boldsymbol v=\boldsymbol w_{t+1}/\|\boldsymbol w_{t+1}\|_2\in\mathbb C^{N\times1}$ denotes the unit-norm beamforming direction. Accordingly, the power constraint becomes $0<\rho\leq P_{\mathrm A}$, with $\|\boldsymbol v\|_2=1$.}
The power and direction can then be determined separately from the structure of the objective. To expose that structure, combine the conditioning terms into the residual channel covariance. Given $\mathbf U$ and $\boldsymbol w_t$, define
\begin{align}
	\mathbf T&\triangleq\alpha^2\bar{\mathbf R}(\mathbf U)\boldsymbol w_t
	\boldsymbol w_t^{\mathsf H}\bar{\mathbf R}(\mathbf U),\label{eq:T_def}\\
	\mathbf H&\triangleq{\bar{\mathbf R}(\mathbf U)-\frac{K_{\mathrm b}}D\mathbf T}.
	\label{eq:conditional_channel_matrix}
\end{align}
The Gaussian conditional-covariance formula gives
$\mathbf H=\operatorname{Cov}(\boldsymbol h_{t+1}(\mathbf U)\mid A_t,B_t)\succeq\mathbf0$.
Equivalently, writing $r_{\mathrm{prev}}=\|\boldsymbol w_t\|_2^2\sigma_z^2/(P_{\mathrm B}+\|\boldsymbol w_t\|_2^2)$ gives $K_{\mathrm b}/D=1/(p+r_{\mathrm{prev}})$. The covariance Cauchy--Schwarz inequality then gives $\boldsymbol v^{\mathsf H}\mathbf H\boldsymbol v\geq[1-\alpha^2p/(p+r_{\mathrm{prev}})]\boldsymbol v^{\mathsf H}\bar{\mathbf R}\boldsymbol v\geq0$ for every unit-norm direction $\boldsymbol v$.
Thus, with $s=\boldsymbol w_{t+1}^{\mathsf H}\mathbf H\boldsymbol w_{t+1}$,
\eqref{eq:N_def}--\eqref{eq:D2_def} become
\begin{align}
	\mathcal N&=\sqrt{P_{\mathrm B}}s,\notag\\
	D_1&=P_{\mathrm B}s+\|\boldsymbol w_{t+1}\|_2^2\sigma_z^2,\notag\\
	D_2&=s+\sigma_z^2.
	\label{eq:beam_residual_components}
\end{align}
{Substitution into~\eqref{eq:X_def} yields}
\begin{align}
	X=\frac{P_{\mathrm B}s^2}
	{(P_{\mathrm B}s+\|\boldsymbol w_{t+1}\|_2^2\sigma_z^2)(s+\sigma_z^2)}.
	\label{eq:beam_residual_objective}
\end{align}
With the power--direction decomposition above, $s=\rho h$, where {$h=\boldsymbol v^{\mathsf H}\mathbf H\boldsymbol v$ is the residual channel power per unit transmit power}. The following proposition determines the optimal power and direction.

\begin{proposition}[Optimal beamforming for a given layout]\label{prop:beam_optimal}
	Suppose that $P_{\mathrm A},P_{\mathrm B},\sigma_z^2>0$ and $\lambda_{\max}(\mathbf H)>0$. An optimal solution of~\eqref{eq:sub1} is
	\begin{align}
		{\boldsymbol w_{t+1}^{\star}
			=\sqrt{P_{\mathrm A}}\,\boldsymbol v_{\max}(\mathbf H)},
		\label{eq:beam_eigen_solution}
	\end{align}
	where {$\boldsymbol v_{\max}(\mathbf H)$ is an optimal direction $\boldsymbol v$, chosen as any unit-norm principal eigenvector of $\mathbf H$}.
\end{proposition}
\begin{proof}
	For a direction with $h>0$, substituting $s=\rho h$ into~\eqref{eq:beam_residual_objective} gives
	\begin{align}
		X(\rho,h)
		=\frac{P_{\mathrm B}\rho h^2}
		{(P_{\mathrm B}h+\sigma_z^2)(\rho h+\sigma_z^2)}.
		\label{eq:beam_power_monotonic}
	\end{align}
	For a given direction, $\rho/(\rho h+\sigma_z^2)$ increases strictly with $\rho$, so the power constraint is active. At $\rho=P_{\mathrm A}$,
	\begin{align}
		\frac{\mathrm d\ln X}{\mathrm ds}
		&=\frac{P_{\mathrm A}\sigma_z^2}
		{s(P_{\mathrm B}s+P_{\mathrm A}\sigma_z^2)}
		+\frac{\sigma_z^2}{s(s+\sigma_z^2)}>0.
		\label{eq:beam_scalar_monotonic}
	\end{align}
	Hence, the objective is maximized by maximizing $s$, giving the equivalent problem
	\begin{align}
		\max_{\|\boldsymbol v\|_2=1}\quad
		\boldsymbol v^{\mathsf H}\mathbf H\boldsymbol v.
		\label{eq:beam_rayleigh_problem}
	\end{align}
	The Rayleigh-quotient bound is attained by a principal eigenvector, proving~\eqref{eq:beam_eigen_solution}. Directions with $h=0$ provide zero information.
\end{proof}

{
For given antenna positions and preceding-slot beamforming, the optimal current-slot beamforming vector can be directly obtained by performing eigenvalue decomposition of $\mathbf H$ and selecting its principal eigenvector. This solution aligns the beamforming direction with the
dominant eigenmode of the conditional channel covariance rather than the unconditioned covariance $\bar{\mathbf R}(\mathbf U)$.
When $\alpha=0$, $\mathbf H=\bar{\mathbf R}(\mathbf U)$, and the conventional principal-eigenvector solution is recovered. If $\mathbf H=\mathbf 0$, the conditional MI is zero for any feasible beamforming vector.
}
%
\endgroup

\subsection{Antenna Position Optimization}

For a given beamforming vector $\boldsymbol w_{t+1}$,
the antenna position optimization problem can be formulated as
\begin{align}
	\max_{\mathbf U}\quad& f(\mathbf U)\notag\\
	\text{s.t.}\quad&
	\|\mathbf u_i-\mathbf u_j\|_2\geq d_{\min},\quad i<j,\notag\\
	&x_n,y_n\in[0,A],\quad n=1,\ldots,N. \label{eq:sub2}
\end{align}

{To address the nonconvex minimum-spacing constraints in Subproblem~\eqref{eq:sub2}, we employ successive convex approximation (SCA)~\cite{XuMA} to construct affine inner approximations, yielding a convex feasible set. For the nonconvex objective function, we adopt projected gradient ascent to iteratively update the antenna positions over the approximated feasible set. Specifically, at a feasible layout $\mathbf U^{(k)}$, the minimum-spacing constraints are approximated as follows:}

\begin{align}
	\frac{(\mathbf u_i^{(k)}-\mathbf u_j^{(k)})^{\top}(\mathbf u_i-\mathbf u_j)}
	{\|\mathbf u_i^{(k)}-\mathbf u_j^{(k)}\|_2}
	\geq d_{\min},\quad i<j. \label{eq:sca_linearization}
\end{align}

{Let $\mathcal F^{(k)}$ contain the layouts satisfying~\eqref{eq:sca_linearization} and the box constraints in~\eqref{problem:P_c3}. The affine bound is tight at $\mathbf U^{(k)}$ and never exceeds the actual antenna separation. Hence, $\mathcal F^{(k)}$ is a nonempty convex subset of the original feasible set. Taking the logarithm changes the objective without changing its maximizers, giving the local position problem}

\begin{align}
	\max_{\mathbf U}&\quad
	2\ln\mathcal N(\mathbf U)
	-\ln D_1(\mathbf U)-\ln D_2(\mathbf U)\notag\\
	&\text{s.t.}\quad\mathbf U\in\mathcal F^{(k)}
	\label{eq:position_local_problem}
\end{align}

{The objective in~\eqref{eq:position_local_problem} remains nonconcave. A projected gradient step uses its coordinate derivatives for the search direction and a convex projection to enforce the local spacing and box constraints.}
{
	{For $u_n\in\{x_n,y_n\}$, differentiation of $\ell_U(\mathbf U)=\ln f(\mathbf U)$ gives}
	
	\begin{align}
		\frac{\partial\ell_U}{\partial u_n}
		&=\left(\frac{2\sqrt{P_{\mathrm B}}}{\mathcal N}
		-\frac{P_{\mathrm B}}{D_1}-\frac{1}{D_2}\right)a_n\notag\\
		&\quad+\left(-\frac{2K_{\mathrm{ab}}}{\mathcal N}
		+\frac{K_{\mathrm a}}{D_1}+\frac{K_{\mathrm b}}{D_2}\right)
		b_n, \label{eq:position_gradient}
	\end{align}
	{
		Let $a_n=\partial q/\partial u_n$ describe the change in current channel power, whereas $b_n=\partial(c/D)/\partial u_n$ describes the change in the correction due to conditioning on the preceding observations.}
	
	The scalar derivatives in~\eqref{eq:position_gradient} are
	\begin{subequations}
		\begin{align}
			a_n&=\frac{\partial q}{\partial u_n}
			=\boldsymbol w_{t+1}^{\mathsf H}\frac{\partial\bar{\mathbf R}}{\partial u_n}\boldsymbol w_{t+1},\\
			b_n&=\frac{\partial(c/D)}{\partial u_n}
			=\frac{1}{D}\frac{\partial c}{\partial u_n}-\frac{c}{D^2}\frac{\partial D}{\partial u_n}, \label{eq:position_scalar_derivatives}\\
			\frac{\partial c}{\partial u_n}
			&=2\alpha^2\mathrm{Re}\left\{(\boldsymbol w_t^{\mathsf H}\bar{\mathbf R}\boldsymbol w_{t+1})^*
			\boldsymbol w_t^{\mathsf H}\frac{\partial\bar{\mathbf R}}{\partial u_n}\boldsymbol w_{t+1}\right\},\\
			\frac{\partial D}{\partial u_n}
			&=\sigma_z^2(P_{\mathrm B}+\|\boldsymbol w_t\|_2^2)
			\boldsymbol w_t^{\mathsf H}\frac{\partial\bar{\mathbf R}}{\partial u_n}\boldsymbol w_t.
		\end{align}
	\end{subequations}
	where $\bar{\mathbf R}$ denotes $\bar{\mathbf R}(\mathbf U)$. The covariance and field-response derivatives are given in Appendix~\ref{sec:position_derivatives}.
	
	{Based on~\eqref{eq:position_gradient}, the projected gradient step for~\eqref{eq:position_local_problem} {is given by}}
	
	\begin{align}
		\mathbf U^{(k+1)}
		=\operatorname{Proj}_{\mathcal F^{(k)}}\left(\mathbf U^{(k)}+\eta_x\nabla_{\mathbf U}\ln f(\mathbf U^{(k)})\right), \label{eq:position_update}
	\end{align}
	where $\eta_x>0$ is the step size and $\nabla_{\mathbf U}\ln f$ collects the coordinate derivatives in~\eqref{eq:position_gradient} in the same $2\times N$ arrangement as $\mathbf U$. {For any trial position matrix $\mathbf Z\in\mathbb R^{2\times N}$, the projection is defined as}
	\begingroup
	\begin{align}
		\operatorname{Proj}_{\mathcal F^{(k)}}(\mathbf Z)
		\triangleq\operatorname*{arg\,min}_{\mathbf U\in\mathcal F^{(k)}}
		\|\mathbf U-\mathbf Z\|_{\mathrm F}^2. \label{eq:proj_F}
	\end{align}
	\endgroup
	
	{The projection in~\eqref{eq:proj_F} is a convex QP in the $2N$ vectorized antenna coordinates, with $4N$ box inequalities and $N(N-1)/2$ linearized spacing inequalities. After updating $\mathbf U^{(k+1)}$, the affine spacing constraints are rebuilt for the next iteration.}
}

{Algorithm~\ref{alg:AO} alternately updates the antenna positions and current-slot beamforming. For a given $\boldsymbol w_{t+1}$, the antenna positions are updated using feasible projected gradient steps with backtracking, where the spacing constraints are relinearized after each accepted update. Subsequently, $\bar{\mathbf R}(\mathbf U)$, $D$, and $\mathbf H$ are recomputed, and the optimal beamforming vector is obtained directly from the principal eigenvector of $\mathbf H$. The preceding-slot beamforming $\boldsymbol w_t$ remains fixed throughout the optimization. Given $\boldsymbol w_t$, the antenna layout and current-slot beamforming are determined before the two-slot probing procedure, with the optimized layout held fixed across both slots.}
{The feasible position updates and exact beamforming updates yield a nondecreasing and bounded objective sequence, ensuring convergence of the objective values. However, global optimality is not guaranteed due to the nonconvexity of the joint problem.}
\begin{algorithm}[ht]
	\caption{AO with Exact Beamforming and Feasible Position Updates}
	\label{alg:AO}
	\begin{algorithmic}[1]
		\Require Given $\boldsymbol w_t\neq\mathbf0$,
		feasible $\mathbf U^{(0)}$, channel and power parameters,
		$\eta_x>0$, $0<\beta<1$,
		tolerances $\varepsilon_{\mathrm{out}},\varepsilon_U>0$,
		limits $I_{\max},K_U,J_{\max}$.
		
		\State Set $i=0$ and compute
		$\bar{\mathbf R}(\mathbf U^{(0)})$, $D$, and $\mathbf H$.
		\State Set
		$\boldsymbol w_{t+1}^{(0)}
		=\sqrt{P_{\mathrm A}}\boldsymbol v_{\max}(\mathbf H)$.
		
		\Repeat
		\State Set $\mathbf U_{\mathrm{loc}}=\mathbf U^{(i)}$
		and evaluate $f_i$.
		
		\For{$k=1,\ldots,K_U$}
		\State Evaluate
		$\boldsymbol G=\nabla_{\mathbf U}\ln f$
		via~\eqref{eq:position_gradient}.
		
		\State Construct $\mathcal F_{\mathrm{loc}}$
		at $\mathbf U_{\mathrm{loc}}$;
		set $\tau=\eta_x$ and
		$\mathbf U_{\mathrm{new}}=\mathbf U_{\mathrm{loc}}$.
		
		\For{$j=1,\ldots,J_{\max}$}
		\State Project
		$\mathbf U_{\mathrm{loc}}+\tau\boldsymbol G$
		onto $\mathcal F_{\mathrm{loc}}$
		via~\eqref{eq:proj_F},
		obtaining $\mathbf U_{\mathrm{trial}}$.
		
		\If{projection succeeds and
			$f(\boldsymbol w_{t+1}^{(i)},\mathbf U_{\mathrm{trial}})
			\geq
			f(\boldsymbol w_{t+1}^{(i)},\mathbf U_{\mathrm{loc}})$}
		\State Set
		$\mathbf U_{\mathrm{new}}=\mathbf U_{\mathrm{trial}}$;
		\textbf{break}.
		\EndIf
		
		\State $\tau\leftarrow\beta\tau$.
		\EndFor
		
		\State Set
		$\Delta_U=
		\|\mathbf U_{\mathrm{new}}-\mathbf U_{\mathrm{loc}}\|_{\mathrm F}$
		and
		$\mathbf U_{\mathrm{loc}}=\mathbf U_{\mathrm{new}}$.
		
		\If{$\Delta_U\leq\varepsilon_U$}
		\State \textbf{break}.
		\EndIf
		\EndFor
		
		\State Set $\mathbf U^{(i+1)}=\mathbf U_{\mathrm{loc}}$.
		
		\State Recompute $\bar{\mathbf R}$, $D$, and $\mathbf H$
		at $\mathbf U^{(i+1)}$.
		
		\State Set
		$\boldsymbol w_{t+1}^{(i+1)}
		=\sqrt{P_{\mathrm A}}\boldsymbol v_{\max}(\mathbf H)$
		and evaluate $f_{i+1}$.
		
		\State Set
		$\Delta_f=|f_{i+1}-f_i|$,
		$\Delta_{\mathrm{AO}}=
		\|\mathbf U^{(i+1)}-\mathbf U^{(i)}\|_{\mathrm F}$,
		and $i\leftarrow i+1$.
		
		\Until{$(\Delta_f\leq\varepsilon_{\mathrm{out}}\
			\textbf{and}\
			\Delta_{\mathrm{AO}}\leq\varepsilon_U)$
			or $i\geq I_{\max}$}
		
		\State \Return
		$\boldsymbol w_{t+1}^{(i)}$ and $\mathbf U^{(i)}$.
	\end{algorithmic}
\end{algorithm}

\subsection{Complexity Analysis}

Constructing $\bar{\mathbf R}(\mathbf U)$ requires
$\mathcal O(N^2N_p)$ operations using dense matrix products.
The rank-one correction in $\mathbf H$ incurs
$\mathcal O(N^2)$ complexity, while the optimal beamforming vector is obtained through Hermitian eigen-decomposition with complexity $\mathcal O(N^3)$.

The position gradient requires $\mathcal O(N^2N_p)$ operations by exploiting the row structure of the field-response derivatives.
Let $C_{\mathrm{QP}}$ denote the complexity of solving the projection problem in~\eqref{eq:proj_F}, which involves $2N$ variables and $4N+N(N-1)/2$ linear inequalities.
With $I_{\mathrm{out}}$ AO iterations, at most $K_U$ position updates per iteration, and at most $J\leq J_{\max}$ backtracking trials per update, the overall computational complexity is
\begin{align}
	\mathcal O\!\left(
	I_{\mathrm{out}}\left[
	N^3+
	K_U\big((J+1)N^2N_p+JC_{\mathrm{QP}}\big)
	\right]\right).
	\label{eq:ao_complexity}
\end{align}
For fixed problem dimensions and iteration limits,
the computational complexity of each two-slot design is independent of the observation duration $T$.

\section{Simulation Results}\label{sec:simulation}
In this section, numerical simulations are conducted to evaluate the performance of the proposed MA-assisted PLKG scheme under the Markov channel model.\footnote{{Unless otherwise stated, the SKR refers to the one-step conditional MI in the simulation.}}
Unless otherwise specified, the system parameters are set as follows\cite{ModelingMA23,ZhuMA23}: $N = 4$ MAs, $N_p = 8$ propagation paths, carrier frequency $f_c = 3$~GHz ($\lambda = 0.1$~m), $d_{\min} = \lambda/2$, movable region $A = 20\lambda$, {$P_{\mathrm{A}} = P_{\mathrm{B}} = 1$~W}, and noise variance $\sigma_z^2 = 0.1$.
The elevation AoDs $\{\theta_p\}_{p=1}^{N_p}$ and azimuth AoDs $\{\phi_p\}_{p=1}^{N_p}$ are uniformly distributed over $[0, \pi]$, and the path gain variance is set to $\sigma_b^2 = 1$.
The temporal correlation coefficient $\alpha$ is varied from $0.1$ to $0.9$ to examine its impact on the SKR.

{Four} representative schemes are evaluated for comparison:
\begin{enumerate}
	\item \textbf{FPA/no opt w}: Fixed-position antennas with no beamforming update, i.e., $\boldsymbol{w}_{t+1} = \boldsymbol{w}_t$. This represents the baseline without any optimization.
	\item \textbf{MA/no opt w}: Movable antennas with position optimization via PGD but fixed beamforming vector $\boldsymbol{w}_{t+1} = \boldsymbol{w}_t$. This isolates the spatial adaptation gain.
	\item \textbf{MA/opt w (Proposed)}: Joint MA position and per-slot beamforming optimization under the AO framework in Algorithm~\ref{alg:AO}, representing the full proposed scheme.
	\item \textbf{Random MA/opt w}: Movable antennas whose positions are drawn uniformly at random within the movable region \cite{Matsumine25} {with the exact beamforming update in~\eqref{eq:beam_eigen_solution}. The randomly selected layouts satisfy the region and minimum-spacing constraints.} 
\end{enumerate}

\subsection{Mutual Information Comparison}

\begin{figure}[htb]
	\centering
	\includegraphics[width=\columnwidth]{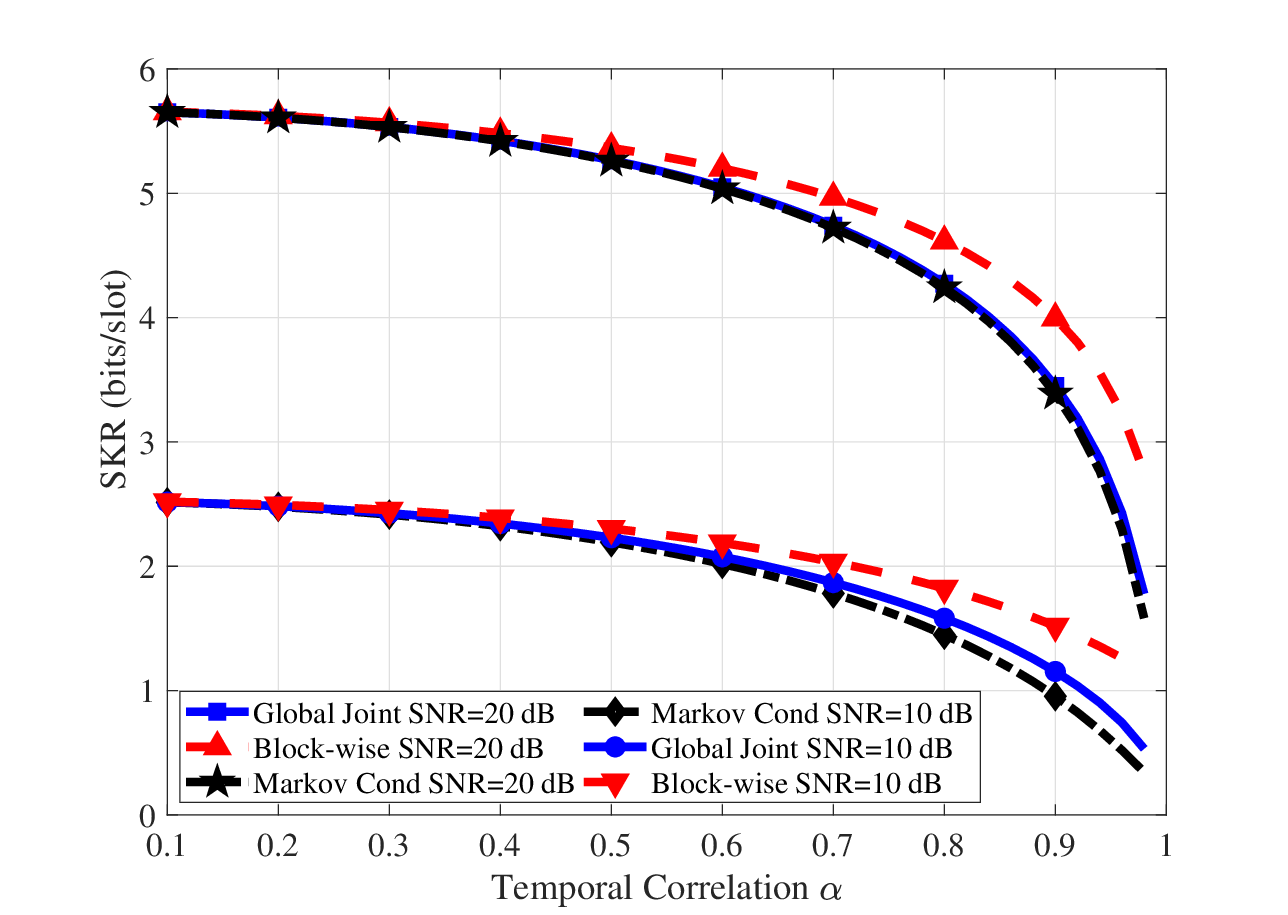}
	\caption{{Comparison of the three MI formulations at SNRs of 10 and 20~dB.}}
	\label{model_comparison_simulation}
\end{figure}
Fig.~\ref{model_comparison_simulation} compares the finite-horizon joint MI $R_T^{\mathrm{joint}}$, the block-wise approximation $R_{T,L}^{\mathrm{block}}$, and the averaged one-step conditional MI $\bar R_T^{(1)}$ at SNRs of 10 and 20~dB with $T=200$. {The antenna positions and beamforming are fixed over all $T$ slots, and the three MI formulations are evaluated using the same configuration.} At both SNRs, the three quantities decrease with $\alpha$, while the approximation discrepancies increase. Stronger temporal correlation makes the dependence discarded across block boundaries more significant. For the one-step formulation, Theorem~\ref{thm:finite_horizon_approx} identifies the cross-terminal predictive information and older-history terms omitted from the full-sequence expansion. The increase of $g_\alpha(\gamma)$ with $\alpha$ at fixed effective SNR is consistent with the widening discrepancy. In contrast, {the smaller discrepancy at 20~dB supports the high-SNR trend established in Theorem~\ref{thm:finite_horizon_approx}}. The theorem concerns the finite-horizon approximation in~\eqref{eq:finite_horizon_error}, whereas the figure plots $\bar R_T^{(1)}$ directly.

\subsection{Impact of Time Correlation}
\begin{figure}[htb]
	\centering
	\includegraphics[width=\columnwidth]{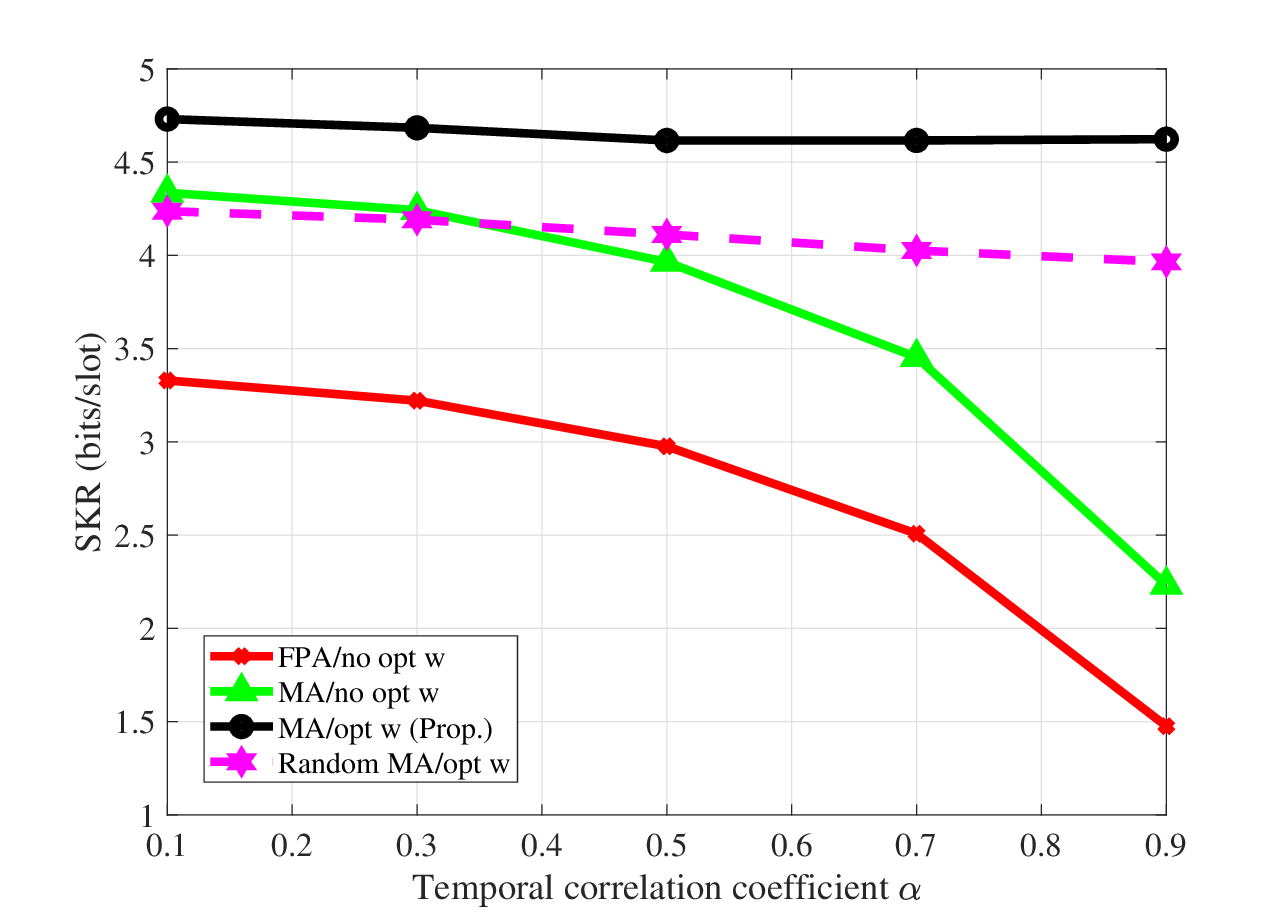}
	\caption{Secret key rate versus the temporal correlation coefficient $\alpha$.}
	\label{Secret key rate versus the temporal correlation}
\end{figure}
Fig.~\ref{Secret key rate versus the temporal correlation} presents the rates of the four schemes as $\alpha$ increases from 0.1 to 0.9. The proposed MA/opt w scheme achieves the highest rate throughout the considered range. The two schemes with unchanged beamforming decrease with increasing $\alpha$, whereas {the proposed scheme varies only slightly over the high-correlation range}. At $\alpha=0.9$, FPA/no opt w, MA/no opt w, and MA/opt w achieve approximately 1.48, 2.23, and 4.62 bits/slot, respectively. Position optimization provides approximately a {$51\%$} gain over the FPA baseline, and joint optimization provides a further {$106\%$} gain over position optimization alone.

The differences follow from the current channel power and the cross-slot correction in~\eqref{eq:N_def}--\eqref{eq:D2_def}. With unchanged beamforming, stronger correlation increases the component predictable from the preceding probe and reduces the residual shared information. Position optimization reshapes $\bar{\mathbf R}(\mathbf U)$ to improve the observation of the available channel variation. Joint design additionally adjusts the next-slot beamforming according to the conditional covariance $\mathbf H$ in~\eqref{eq:conditional_channel_matrix}, balancing current observation quality against dependence on the preceding pair. {These gains reflect more effective use of the available channel randomness through spatial design, while the temporal evolution of the path gains follows the prescribed Gauss--Markov model.}

The crossover between MA/no opt w and Random MA/opt w further illustrates the role of beamforming. At low $\alpha$, the conditioning correction is weak, and an optimized layout with unchanged beamforming outperforms a random layout with optimized beamforming. At higher $\alpha$, next-slot beamforming adaptation becomes more beneficial, allowing Random MA/opt w to exceed MA/no opt w. Combining layout and beamforming adaptation accounts for the advantage of the proposed scheme over both alternatives.

\subsection{Impact of the Number of Movable Antennas}
\begin{figure}[htb]
	\centering
	\includegraphics[width=\columnwidth]{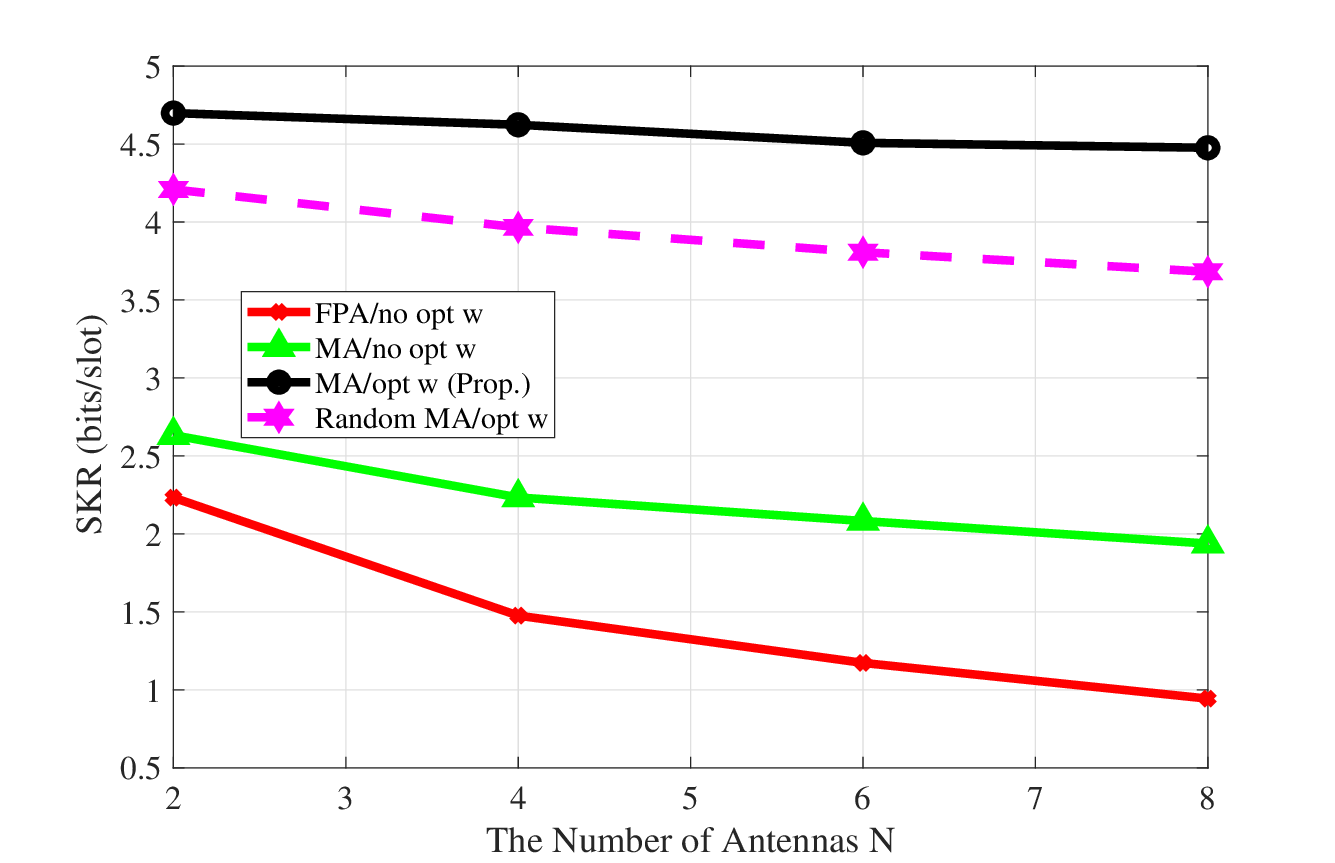}
	\caption{Secret key rate versus the number of movable antennas $N$.}
	\label{Secret key rate versus the number of movable antenna}
\end{figure}
Fig.~\ref{Secret key rate versus the number of movable antenna} shows the rate versus $N$ at $\alpha=0.9$. The proposed scheme maintains the highest rate and exhibits only a modest decrease over $N=2$ to 8, while the FPA/no opt w baseline decreases more noticeably. MA/no opt w remains above the FPA baseline, demonstrating the benefit of position optimization even without a beamforming update.

The antenna-number dependence should be interpreted under the {unit-norm steering-vector normalization} in~\eqref{eq:steering_2d}. As stated in Remark~1, $\operatorname{tr}(\bar{\mathbf R})=N_p\sigma_b^2$ remains constant. Increasing $N$ does not automatically increase total channel energy. With a fixed power budget, a nonadaptive beamforming can capture a smaller share of that energy, and the particular layout also affects the effective channel power. Joint optimization selects the antenna geometry and spatial weighting through $\mathbf H$, retaining a higher conditional rate under the adopted normalization.

\subsection{Impact of Alice Power Budget}
\begin{figure}[htb]
	\centering
	\includegraphics[width=\columnwidth]{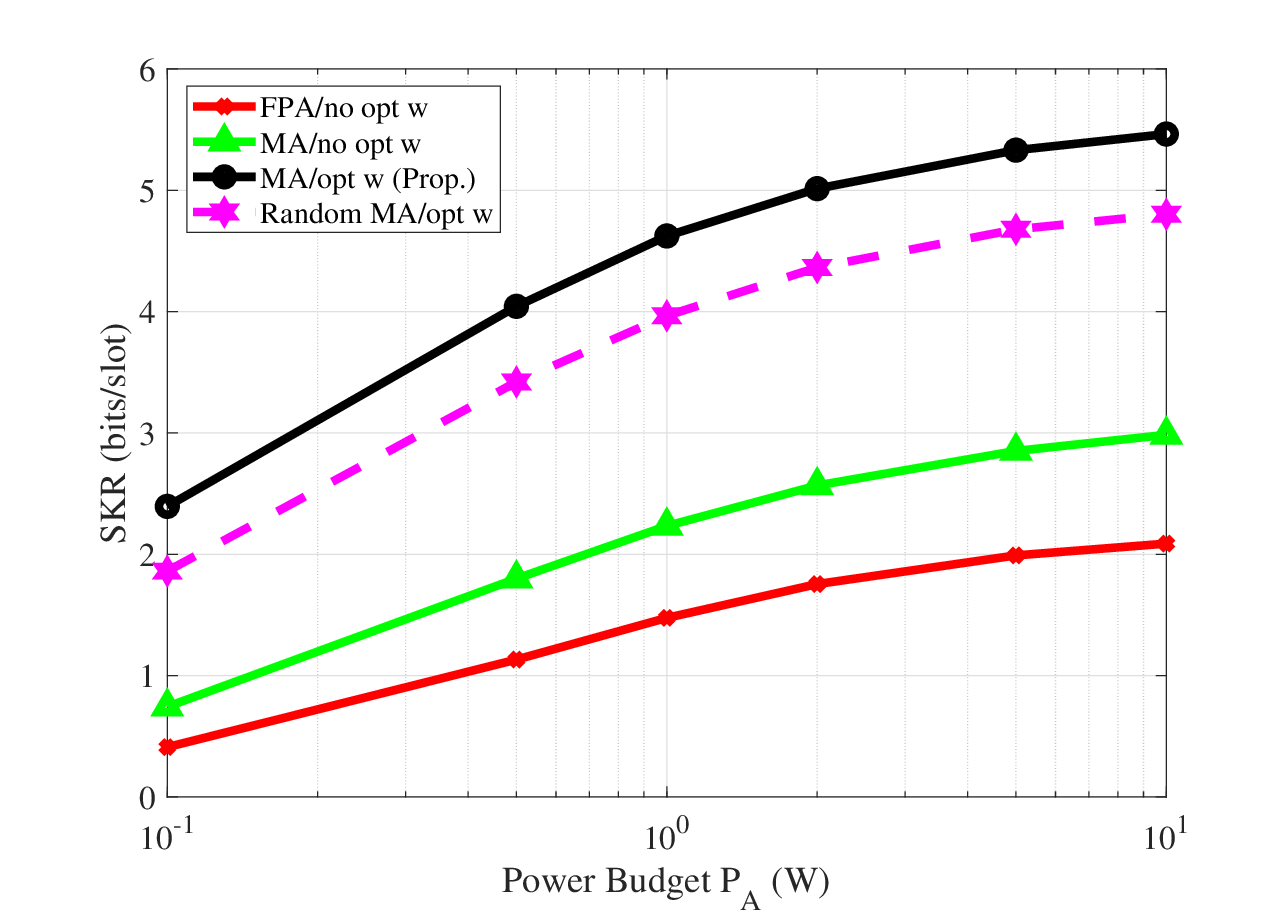}
	\caption{Secret key rate versus Alice transmit power budget $P_{\mathrm{A}}$.}
	\label{Secret key rate versus Alice power budget}
\end{figure}
\begin{figure*}[htb]
	\centering
	\includegraphics[width=0.33\textwidth]{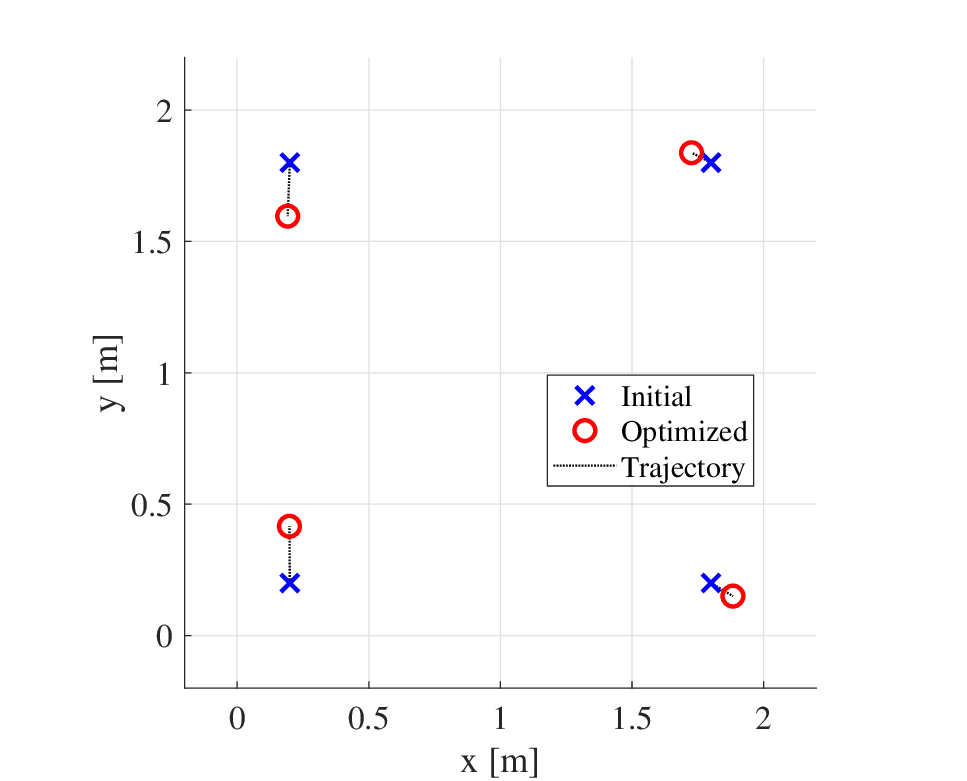}\hfill
	\includegraphics[width=0.33\textwidth]{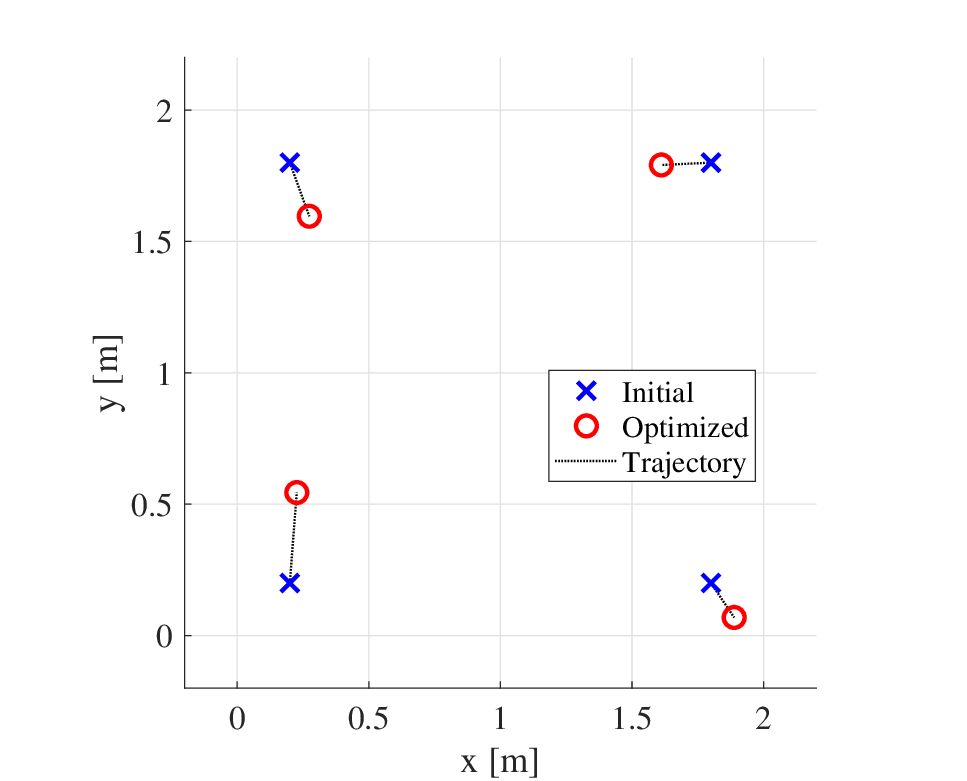}\hfill
	\includegraphics[width=0.33\textwidth]{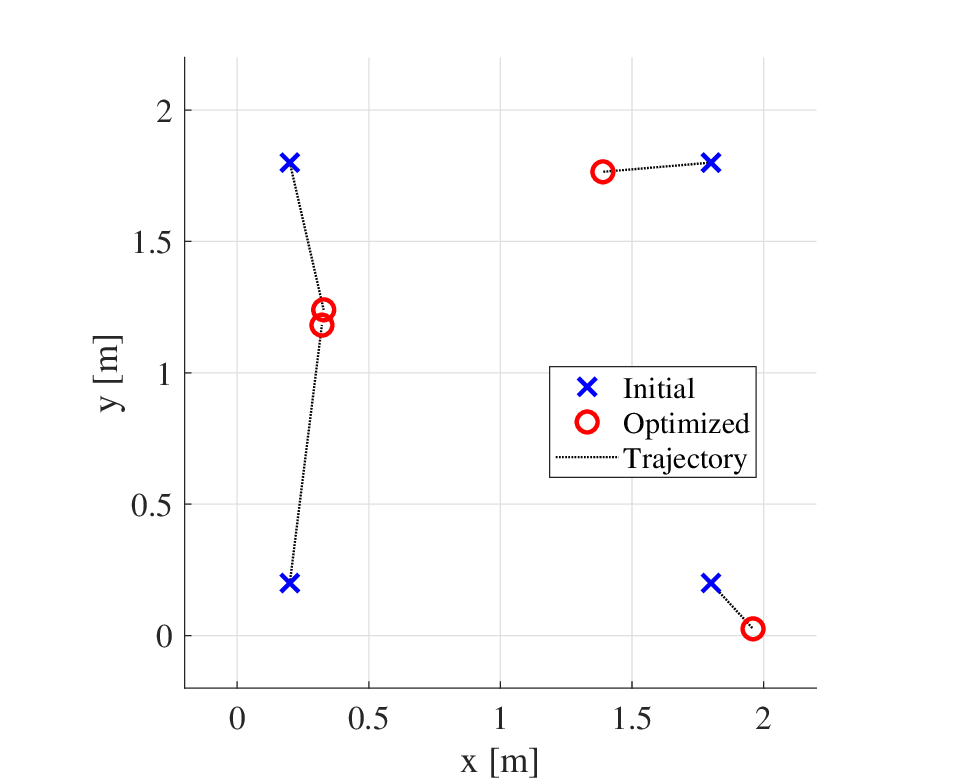}
	\caption{{Initial and optimized antenna positions for (a) $\alpha=0.1$, (b) $\alpha=0.5$, and (c) $\alpha=0.9$.}}\label{ma_trajectory}
\end{figure*}
Fig.~\ref{Secret key rate versus Alice power budget} presents the rate versus $P_{\mathrm A}$ at $\alpha=0.9$. All schemes improve as the power budget increases, and the proposed joint design consistently achieves the highest rate. The separation between MA/opt w and MA/no opt w becomes more pronounced over the considered range, showing the additional benefit of beamforming adaptation beyond position optimization.

The power dependence agrees with the beam-scaling analysis in Proposition~\ref{prop:beam_optimal}. Increasing the beam norm scales Alice's effective signal and combined-noise powers together, while Bob's receiver-noise variance remains unchanged. Hence, {increasing $P_{\mathrm A}$ improves Bob's observation quality without reducing Alice's effective SNR for a given beam direction}. Position and beamforming optimization further increase the residual channel power entering~\eqref{eq:beam_residual_objective}. The rate curves reflect the combined effects of transmit power and spatial design, with the latter providing higher rates at the same power budget.

\subsection{Movable Antenna Deployment}

Fig.~\ref{ma_trajectory} compares the initial and optimized antenna positions for $\alpha=0.1$, 0.5, and 0.9. All three cases start from a uniform $2\times2$ grid within the same region $\mathcal C=[0,A]\times[0,A]$. In this numerical example, the initial-to-final displacement of each antenna increases as $\alpha$ rises from 0.1 to 0.9, indicating more pronounced position adjustments under stronger temporal correlation.

The connecting lines indicate initial-to-final displacements rather than movement during probing. The selected positions remain fixed over the analyzed probing interval. Throughout the numerical updates, the projection in~\eqref{eq:position_update} enforces the region and minimum-spacing constraints. The configurations illustrate spatial adaptation within the available region, without requiring a larger movement area.

\section{Conclusion}\label{sec:conclusion}
{PLKG enables wireless key distribution through reciprocal channel observations, but strong temporal dependence limits the fresh randomness available in slowly varying channels. This paper investigated MA-assisted PLKG with Gauss--Markov multipath gains. A closed-form one-step conditional MI was derived. {For fixed antenna positions and beamforming, the finite-horizon approximation error was shown to vanish as both effective SNRs increased.} Joint antenna-position and per-slot beamforming design was addressed by AO, combining an exact {principal-eigenvector beamforming solution} with feasible SCA-based projected gradient position updates. The simulations demonstrated improvements over the considered fixed-position baseline and additional gains from beamforming adaptation. At $\alpha=0.9$, the plotted results {indicated} approximately a {$51\%$} improvement from position optimization and a further {$106\%$} improvement from joint design, supporting the benefit of spatial adaptation under strong temporal correlation.}

\appendix
\begin{appendices}
	{
		\subsection{Proof of Theorem~\ref{thm:finite_horizon_approx}}\label{app:finite_horizon_approx}
		
		The finite-horizon approximation bound and high-SNR consistency in Theorem~\ref{thm:finite_horizon_approx} are proved below.
		
		Set $Y_t=(A_t,B_t)$, $A^t=(A_1,\ldots,A_t)$, $B^t=(B_1,\ldots,B_t)$, and $Y^t=(Y_1,\ldots,Y_t)$. For $t\geq2$, define
		\begin{align}
			\delta_{\mathrm a,t}&=\mathcal I(A_t;B^{t-1}\mid A^{t-1}),\notag\\
			\delta_{\mathrm b,t}&=\mathcal I(B_t;A^{t-1}\mid B^{t-1}),\notag\\
			\eta_t&=\mathcal I(Y_t;Y^{t-2}\mid Y_{t-1}), \label{eq:error_terms_app}
		\end{align}
		with $\eta_2=0$. The first two terms represent cross-terminal predictive information, while $\eta_t$ controls the effect of truncating the observation history. For finite mutual informations, the error in~\eqref{eq:finite_horizon_error} obeys the general bound
		\begin{align}
			\mathcal E_T\leq\frac1T\sum_{t=2}^T
			(\delta_{\mathrm a,t}+\delta_{\mathrm b,t}+\eta_t). \label{eq:one_step_error_bound}
		\end{align}
		Let $J_t=\mathcal{I}(A^t;B^t)$. The mutual-information chain rule~\cite[Ch.~2]{CoverThomas2006} gives
		\begin{align}
			J_t
			&=\mathcal{I}(A^{t-1};B^{t-1},B_t)\notag\\
			&\quad+\mathcal{I}(A_t;B^{t-1},B_t\mid A^{t-1})\notag\\
			&=J_{t-1}+\mathcal{I}(B_t;A^{t-1}\mid B^{t-1})\notag\\
			&\quad+\mathcal{I}(A_t;B^{t-1}\mid A^{t-1})\notag\\
			&\quad+\mathcal{I}(A_t;B_t\mid Y^{t-1}). \label{eq:exact_mi_increment_app}
		\end{align}
		Thus, the increment contains the two cross-terminal terms. Next, set $Z_t=Y^{t-2}$, $F_t=\mathcal{I}(A_t;B_t\mid Y_{t-1},Z_t)$, and $M_t=\mathcal{I}(A_t;B_t\mid Y_{t-1})=R_{t-1}^{(1)}$. Two chain-rule expansions give
		\begin{align}
			M_t-F_t
			&=\mathcal{I}(A_t;Z_t\mid Y_{t-1})\notag\\
			&\quad-\mathcal{I}(A_t;Z_t\mid B_t,Y_{t-1}). \label{eq:cmi_difference_app}
		\end{align}
		Moreover,
		\begin{align}
			\eta_t
			&=\mathcal{I}(A_t;Z_t\mid Y_{t-1})+\mathcal{I}(B_t;Z_t\mid A_t,Y_{t-1})\notag\\
			&=\mathcal{I}(B_t;Z_t\mid Y_{t-1})+\mathcal{I}(A_t;Z_t\mid B_t,Y_{t-1}).
			\notag
		\end{align}
		Thus, both terms on the right-hand side of~\eqref{eq:cmi_difference_app} lie in $[0,\eta_t]$, so
		\begin{align}
			|F_t-M_t|\leq\eta_t. \label{eq:cmi_memory_bound_app}
		\end{align}
		At $t=2$, $F_2=M_2$ and $\eta_2=0$. Summing~\eqref{eq:exact_mi_increment_app} gives
		\begin{align}
			&J_T-J_1-\sum_{t=2}^{T}M_t\notag\\
			&\quad=\sum_{t=2}^{T}\big[\delta_{\mathrm{a},t}+\delta_{\mathrm{b},t}+F_t-M_t\big].\notag
		\end{align}
		Applying~\eqref{eq:cmi_memory_bound_app} and the triangle inequality, dividing by $T$, and using $J_1=\mathcal{I}(A_1;B_1)$ and $\sum_{t=2}^{T}M_t=(T-1)\bar R_T^{(1)}$ prove~\eqref{eq:one_step_error_bound}.
		
		For the Gauss--Markov specialization, let $G_t=\mathbf w^{\mathsf H}\mathbf A(\mathbf U)\mathbf b_t$. Then $G_t=\alpha G_{t-1}+V_t$, where $V_t\sim\mathcal{CN}(0,{\sigma_V^2})$ and ${\sigma_V^2}=(1-\alpha^2)S$. {Since mutual information is invariant under invertible scaling, we may, without loss of generality, work with the equivalent normalized observation model $A_t=G_t+N_{\mathrm a,t}$ and $B_t=G_t+N_{\mathrm b,t}$, where the noise variances are $r_{\mathrm a}$ and $r_{\mathrm b}$, respectively.}
		
		For jointly proper Gaussian variables, the conditional entropy formula~\cite{CoverThomas2006} expresses $\delta_{\mathrm a,t}$ as a conditional-variance ratio. To bound this ratio, it is sufficient to compare the best prediction based on one terminal's history with a simple feasible predictor. For any observation history $\mathcal H$, the conditional mean minimizes the mean-square prediction error. Indeed, for any square-integrable predictor $f(\mathcal H)$,
		\begin{align}
			\mathbb E|A_t-f(\mathcal H)|^2
			&=\mathbb E|A_t-\mathbb E[A_t\mid\mathcal H]|^2\notag\\
			&\quad+\mathbb E|\mathbb E[A_t\mid\mathcal H]-f(\mathcal H)|^2. \label{eq:prediction_orthogonality_app}
		\end{align}
		The cross term vanishes because the prediction residual has zero conditional mean given $\mathcal H$. For jointly Gaussian observations, the conditional variance does not depend on the realized history, so it equals the minimum mean-square prediction error.
		
		Since $V_t$ and $N_{\mathrm a,t}$ are independent of both terminals' past observations, conditioning on any past history leaves their variances unchanged. In particular,
		\begin{align}
			\operatorname{Var}(A_t\mid A^{t-1},B^{t-1})
			&=\alpha^2\operatorname{Var}(G_{t-1}\mid A^{t-1},B^{t-1})\notag\\
			&\quad+{\sigma_V^2}+r_{\mathrm a}\geq {\sigma_V^2}+r_{\mathrm a}. \label{eq:innovation_prediction_floor_app}
		\end{align}
		Here ${\sigma_V^2}$ is the power of the new channel-state innovation, not an increment caused by receiver noise. The two contributions ${\sigma_V^2}$ and $r_{\mathrm a}$ are independent and both remain unpredictable from earlier probes.
		
		For the numerator, the predictor $\alpha A_{t-1}$ uses only information available in $A^{t-1}$ and has error $V_t+N_{\mathrm a,t}-\alpha N_{\mathrm a,t-1}$. Its mean-square error is ${\sigma_V^2}+r_{\mathrm a}+\alpha^2r_{\mathrm a}$. The optimal predictor based on the entire history $A^{t-1}$ cannot have a larger error, yielding $\operatorname{Var}(A_t\mid A^{t-1})\leq {\sigma_V^2}+r_{\mathrm a}+\alpha^2r_{\mathrm a}$. Combining this upper bound with~\eqref{eq:innovation_prediction_floor_app} gives
		\begin{align}
			\delta_{\mathrm a,t}
			&=\log_2\frac{\operatorname{Var}(A_t\mid A^{t-1})}
			{\operatorname{Var}(A_t\mid A^{t-1},B^{t-1})}\notag\\
			&\leq\log_2\!\left(1+\frac{\alpha^2r_{\mathrm a}}{{\sigma_V^2}+r_{\mathrm a}}\right)\notag\\
			&=\log_2\!\left(1+\frac{\alpha^2}{1+(1-\alpha^2)\gamma_{\mathrm a}}\right)
			=g_\alpha(\gamma_{\mathrm a}). \label{eq:delta_a_snr_bound_app}
		\end{align}
		By symmetry, $\delta_{\mathrm b,t}\leq g_\alpha(\gamma_{\mathrm b})$.
		
		For $r_{\mathrm{ab}}=(r_{\mathrm a}^{-1}+r_{\mathrm b}^{-1})^{-1}$, define the equivalent observation
		\begin{align}
			C_t=r_{\mathrm{ab}}\left(\frac{A_t}{r_{\mathrm a}}+\frac{B_t}{r_{\mathrm b}}\right)=G_t+N_{\mathrm{ab},t}, \label{eq:combined_obs_app}
		\end{align}
		where $N_{\mathrm{ab},t}$ has variance $r_{\mathrm{ab}}$. The weights give more emphasis to the terminal with the smaller observation-noise variance. This combination is used only in the proof to represent the information in an observation pair, not as an additional terminal-side processing requirement.
		
		Let $D_t^{\mathrm n}=A_t-B_t=N_{\mathrm a,t}-N_{\mathrm b,t}$. The transformation from $(A_t,B_t)$ to $(C_t,D_t^{\mathrm n})$ is invertible, and
		\begin{align}
			\mathbb E\{N_{\mathrm{ab},t}(D_t^{\mathrm n})^*\}
			=r_{\mathrm{ab}}\left(\frac{r_{\mathrm a}}{r_{\mathrm a}}
			-\frac{r_{\mathrm b}}{r_{\mathrm b}}\right)=0. \label{eq:noise_complement_app}
		\end{align}
		Joint Gaussianity turns this zero covariance into independence. Moreover, $D_t^{\mathrm n}$ is independent of the channel and all noises at other slots. Thus, the complementary noise sequence carries no temporal information, giving $\eta_t=\mathcal I(C_t;C^{t-2}\mid C_{t-1})$.
		
		The conditional variance given the full preceding history is at least ${\sigma_V^2}+r_{\mathrm{ab}}$, since the current innovation and combined noise are independent of that history. With only $C_{t-1}$ available, the predictor $\alpha C_{t-1}$ has error variance ${\sigma_V^2}+r_{\mathrm{ab}}+\alpha^2r_{\mathrm{ab}}$. The Gaussian conditional-entropy ratio thus yields
		\begin{align}
			\eta_t=\mathcal I(C_t;C^{t-2}\mid C_{t-1})
			\leq g_\alpha\!\left(\frac{S}{r_{\mathrm{ab}}}\right)
			=g_\alpha(\gamma_{\mathrm a}+\gamma_{\mathrm b}), \label{eq:eta_snr_bound_app}
		\end{align}
		because $S/r_{\mathrm{ab}}=\gamma_{\mathrm a}+\gamma_{\mathrm b}$. Substitution into~\eqref{eq:one_step_error_bound} proves~\eqref{eq:high_snr_error_bound} and~\eqref{eq:high_snr_limit}.
	}
	\subsection{Covariance Matrix Definitions}\label{sec:cov_matrices}
	
	For notational simplicity, let
	$\hat h_{\mathrm a}=A_{t+1}$,
	$\hat h_{\mathrm b}=B_{t+1}$,
	$\hat h_{\mathrm a'}=A_t$, and
	$\hat h_{\mathrm b'}=B_t$.
	Define
	$\mathcal R_{\mathrm{xy}}
	=\mathbb E\{\hat h_{\mathrm x}\hat h_{\mathrm y}^*\}$
	for $\mathrm x,\mathrm y
	\in\{\mathrm a,\mathrm b,\mathrm a',\mathrm b'\}$.
	The covariance matrices in~\eqref{eq:det_formula}
	are given by

	\begin{align}
		\mathbf{R}_{\mathrm{a}'\mathrm{b}'} &= \begin{bmatrix}
			\mathcal{R}_{\mathrm{a}'\mathrm{a}'} & \mathcal{R}_{\mathrm{a}'\mathrm{b}'} \\
			\mathcal{R}_{\mathrm{b}'\mathrm{a}'} & \mathcal{R}_{\mathrm{b}'\mathrm{b}'}
		\end{bmatrix}, \label{eq:R_a'b'_app} \\
		\mathbf{R}_{\mathrm{aa}'\mathrm{b}'} &= \begin{bmatrix}
			\mathcal{R}_{\mathrm{aa}} & \mathcal{R}_{\mathrm{aa}'} & \mathcal{R}_{\mathrm{ab}'} \\
			\mathcal{R}_{\mathrm{a}'\mathrm{a}} & \mathcal{R}_{\mathrm{a}'\mathrm{a}'} & \mathcal{R}_{\mathrm{a}'\mathrm{b}'} \\
			\mathcal{R}_{\mathrm{b}'\mathrm{a}} & \mathcal{R}_{\mathrm{b}'\mathrm{a}'} & \mathcal{R}_{\mathrm{b}'\mathrm{b}'}
		\end{bmatrix}, \label{eq:R_aa'b'_app} \\
		\mathbf{R}_{\mathrm{ba}'\mathrm{b}'} &= \begin{bmatrix}
			\mathcal{R}_{\mathrm{bb}} & \mathcal{R}_{\mathrm{ba}'} & \mathcal{R}_{\mathrm{bb}'} \\
			\mathcal{R}_{\mathrm{a}'\mathrm{b}} & \mathcal{R}_{\mathrm{a}'\mathrm{a}'} & \mathcal{R}_{\mathrm{a}'\mathrm{b}'} \\
			\mathcal{R}_{\mathrm{b}'\mathrm{b}} & \mathcal{R}_{\mathrm{b}'\mathrm{a}'} & \mathcal{R}_{\mathrm{b}'\mathrm{b}'}
		\end{bmatrix}, \label{eq:R_ba'b'_app} \\
		\mathbf{R}_{\mathrm{aba}'\mathrm{b}'} &= \begin{bmatrix}
			\mathcal{R}_{\mathrm{aa}} & \mathcal{R}_{\mathrm{ab}} & \mathcal{R}_{\mathrm{aa}'} & \mathcal{R}_{\mathrm{ab}'} \\
			\mathcal{R}_{\mathrm{ba}} & \mathcal{R}_{\mathrm{bb}} & \mathcal{R}_{\mathrm{ba}'} & \mathcal{R}_{\mathrm{bb}'} \\
			\mathcal{R}_{\mathrm{a}'\mathrm{a}} & \mathcal{R}_{\mathrm{a}'\mathrm{b}} & \mathcal{R}_{\mathrm{a}'\mathrm{a}'} & \mathcal{R}_{\mathrm{a}'\mathrm{b}'} \\
			\mathcal{R}_{\mathrm{b}'\mathrm{a}} & \mathcal{R}_{\mathrm{b}'\mathrm{b}} & \mathcal{R}_{\mathrm{b}'\mathrm{a}'} & \mathcal{R}_{\mathrm{b}'\mathrm{b}'}
		\end{bmatrix}, \label{eq:R_aba'b'_app}
	\end{align}

	\subsection{Detailed Covariance Derivations}\label{sec:cov_derivations}
	
	For completeness, the {same-slot} covariance matrix $\bar{\mathbf{R}}(\mathbf{U})$ in \eqref{eq:bar_R} and the {adjacent-slot} covariance matrix $\tilde{\mathbf{R}}(\mathbf{U})$ in \eqref{eq:tilde_R} are obtained under the MA model as follows.
	For the channel vector $\mathbf{A}(\mathbf{U}) \boldsymbol{b}_t$ with {a deterministic field-response matrix $\mathbf{A}(\mathbf{U})$ and random path gains}, substituting $\mathbb{E}\{\boldsymbol{b}_t \boldsymbol{b}_t^{\mathsf{H}}\} = \sigma_b^2 \mathbf{I}_{N_p}$ and the Markov cross-correlation $\mathbb{E}\{\boldsymbol{b}_t \boldsymbol{b}_{t+1}^{\mathsf{H}}\} = \alpha \sigma_b^2 \mathbf{I}_{N_p}$ {yields}
	\begin{align}
		\bar{\mathbf{R}}(\mathbf{U}) &= \sigma_b^2 \mathbf{A}(\mathbf{U}) \mathbf{A}(\mathbf{U})^{\mathsf{H}}, \\
		\tilde{\mathbf{R}}(\mathbf{U}) &= \alpha \sigma_b^2 \mathbf{A}(\mathbf{U}) \mathbf{A}(\mathbf{U})^{\mathsf{H}} = \alpha \bar{\mathbf{R}}(\mathbf{U}), \label{eq:tilde_R}
	\end{align}
	which establishes \eqref{eq:bar_R} and \eqref{eq:tilde_R}.
	\subsection{Covariance Elements}\label{sec:cov_elements}
	With $\bar{\mathbf{R}}(\mathbf{U})$ defined in \eqref{eq:bar_R} and $\tilde{\mathbf{R}}(\mathbf{U}) = \alpha \bar{\mathbf{R}}(\mathbf{U})$ defined in \eqref{eq:tilde_R}, and assuming uniform noise variance $\sigma_z^2$, the scalar covariance elements are given by:
	\begin{align}
		\mathcal{R}_{\mathrm{aa}} &= \mathbb{E}\{\hat{h}_{\mathrm{a}} \hat{h}_{\mathrm{a}}^{\mathsf{H}}\}
		= P_{\mathrm{B}} \boldsymbol{w}_{t+1}^{\mathsf{H}} \bar{\mathbf{R}}(\mathbf{U}) \boldsymbol{w}_{t+1} + \|\boldsymbol{w}_{t+1}\|_2^2 \sigma_z^2, \label{eq:R_aa_app} \\
		\mathcal{R}_{\mathrm{bb}} &= \mathbb{E}\{\hat{h}_{\mathrm{b}} \hat{h}_{\mathrm{b}}^{\mathsf{H}}\}
		= \boldsymbol{w}_{t+1}^{\mathsf{H}} \bar{\mathbf{R}}(\mathbf{U}) \boldsymbol{w}_{t+1} + \sigma_z^2, \label{eq:R_bb_app} \\
		\mathcal{R}_{\mathrm{a}'\mathrm{a}'} &= \mathbb{E}\{\hat{h}_{\mathrm{a}'} \hat{h}_{\mathrm{a}'}^{\mathsf{H}}\}
		= P_{\mathrm{B}} \boldsymbol{w}_{t}^{\mathsf{H}} \bar{\mathbf{R}}(\mathbf{U}) \boldsymbol{w}_{t} + \|\boldsymbol{w}_{t}\|_2^2 \sigma_z^2, \label{eq:R_a'a'_app} \\
		\mathcal{R}_{\mathrm{b}'\mathrm{b}'} &= \mathbb{E}\{\hat{h}_{\mathrm{b}'} \hat{h}_{\mathrm{b}'}^{\mathsf{H}}\}
		= \boldsymbol{w}_{t}^{\mathsf{H}} \bar{\mathbf{R}}(\mathbf{U}) \boldsymbol{w}_{t} + \sigma_z^2, \label{eq:R_b'b'_app} \\
		\mathcal{R}_{\mathrm{a}'\mathrm{b}'} &= \mathbb{E}\{\hat{h}_{\mathrm{a}'} \hat{h}_{\mathrm{b}'}^{\mathsf{H}}\}
		= \sqrt{P_{\mathrm{B}}}\, \boldsymbol{w}_{t}^{\mathsf{H}} \bar{\mathbf{R}}(\mathbf{U}) \boldsymbol{w}_{t} = \mathcal{R}_{\mathrm{b}'\mathrm{a}'}^{\mathsf{H}}, \label{eq:R_a'b'_cov_app} \\
		\mathcal{R}_{\mathrm{ab}} &= \mathbb{E}\{\hat{h}_{\mathrm{a}} \hat{h}_{\mathrm{b}}^{\mathsf{H}}\}
		= \sqrt{P_{\mathrm{B}}}\, \boldsymbol{w}_{t+1}^{\mathsf{H}} \bar{\mathbf{R}}(\mathbf{U}) \boldsymbol{w}_{t+1} = \mathcal{R}_{\mathrm{ba}}^{\mathsf{H}}, \label{eq:R_ab_app} \\
		\mathcal{R}_{\mathrm{a}\mathrm{b}'} &= \mathbb{E}\{\hat{h}_{\mathrm{a}} \hat{h}_{\mathrm{b}'}^{\mathsf{H}}\}
		= \sqrt{P_{\mathrm{B}}}\, \alpha \, \boldsymbol{w}_{t+1}^{\mathsf{H}} \bar{\mathbf{R}}(\mathbf{U}) \boldsymbol{w}_{t} = \mathcal{R}_{\mathrm{b}'\mathrm{a}}^{\mathsf{H}}, \label{eq:R_ab'_app} \\
		\mathcal{R}_{\mathrm{a}'\mathrm{a}} &= \mathbb{E}\{\hat{h}_{\mathrm{a}'} \hat{h}_{\mathrm{a}}^{\mathsf{H}}\}
		= P_{\mathrm{B}} \alpha \, \boldsymbol{w}_{t}^{\mathsf{H}} \bar{\mathbf{R}}(\mathbf{U}) \boldsymbol{w}_{t+1} = \mathcal{R}_{\mathrm{aa}'}^{\mathsf{H}}, \label{eq:R_a'a_app} \\
		\mathcal{R}_{\mathrm{a}'\mathrm{b}} &= \mathbb{E}\{\hat{h}_{\mathrm{a}'} \hat{h}_{\mathrm{b}}^{\mathsf{H}}\}
		= \sqrt{P_{\mathrm{B}}}\, \alpha \, \boldsymbol{w}_{t}^{\mathsf{H}} \bar{\mathbf{R}}(\mathbf{U}) \boldsymbol{w}_{t+1} = \mathcal{R}_{\mathrm{ba}'}^{\mathsf{H}}, \label{eq:R_a'b_app} \\
		\mathcal{R}_{\mathrm{b}'\mathrm{b}} &= \mathbb{E}\{\hat{h}_{\mathrm{b}'} \hat{h}_{\mathrm{b}}^{\mathsf{H}}\}
		= \alpha \, \boldsymbol{w}_{t}^{\mathsf{H}} \bar{\mathbf{R}}(\mathbf{U}) \boldsymbol{w}_{t+1} = \mathcal{R}_{\mathrm{bb}'}^{\mathsf{H}}. \label{eq:R_b'b_app}
	\end{align}
	

	\subsection{Schur Complement Proofs}\label{sec:schur_proof}
	
	For a block matrix with nonsingular pivot $\mathbf Q$, the Schur complement gives
	\begin{align}
		\det\begin{bmatrix}\mathbf M&\mathbf B\\\mathbf C&\mathbf Q\end{bmatrix}
		=\det(\mathbf Q)\det(\mathbf M-\mathbf B\mathbf Q^{-1}\mathbf C).
		\label{eq:schur}
	\end{align}
	
	The appendix provides the Schur complement derivations of the determinant expressions \eqref{eq:D_def} and~\eqref{eq:det_Raa'b'}--\eqref{eq:det_Raba'b'} obtained from the block determinant identity~\eqref{eq:schur}.
	
	\emph{Determinant of $\mathbf{R}_{\mathrm{a}'\mathrm{b}'}$ }:
	For the block matrix in \eqref{eq:R_a'b'_app}, the determinant follows directly:
	\begin{align}
		D = \det(\mathbf{R}_{\mathrm{a}'\mathrm{b}'}) = \mathcal{R}_{\mathrm{a}'\mathrm{a}'} \mathcal{R}_{\mathrm{b}'\mathrm{b}'} - |\mathcal{R}_{\mathrm{a}'\mathrm{b}'}|^2.
	\end{align}
	{For fixed $\mathbf U$ and a given nonzero $\boldsymbol w_t$, $D$ is a known positive scalar because $\mathbf R_{\mathrm{a}'\mathrm{b}'}$ is positive definite under nonzero noise variance. Its dependence on $\mathbf U$ is retained in the position update.}
	
	\emph{Determinant of $\mathbf{R}_{\mathrm{aa}'\mathrm{b}'}$ and $\mathbf{R}_{\mathrm{ba}'\mathrm{b}'}$}:
	Partitioning each augmented matrix in \eqref{eq:R_aa'b'_app} and \eqref{eq:R_ba'b'_app} as $\begin{bmatrix} \mathcal{R}_{\mathrm{xx}} & \mathbf{B} \\ \mathbf{C} & \mathbf{R}_{\mathrm{a}'\mathrm{b}'} \end{bmatrix}$ with $\mathrm{x} \in \{\mathrm{a}, \mathrm{b}\}$, $\mathbf{B} = [\mathcal{R}_{\mathrm{xa}'}, \mathcal{R}_{\mathrm{xb}'}]$, and $\mathbf{C} = [\mathcal{R}_{\mathrm{a}'\mathrm{x}}; \mathcal{R}_{\mathrm{b}'\mathrm{x}}]$, and applying the Schur complement with respect to the pivot block $\mathbf{R}_{\mathrm{a}'\mathrm{b}'}$ yields
	\begin{align}
		\det(\mathbf{R}_{\mathrm{aa}'\mathrm{b}'}) = D \cdot \big(\mathcal{R}_{\mathrm{aa}} - \mathbf{B} \mathbf{R}_{\mathrm{a}'\mathrm{b}'}^{-1} \mathbf{C}\big).
	\end{align}
	Evaluating the quadratic form $\mathbf{B} \mathbf{R}_{\mathrm{a}'\mathrm{b}'}^{-1} \mathbf{C}$ with the covariance elements in \eqref{eq:R_aa_app}--\eqref{eq:R_b'b_app} and the definitions of $K_{\mathrm{a}}$ yields \eqref{eq:det_Raa'b'}. The expression for $\det(\mathbf{R}_{\mathrm{ba}'\mathrm{b}'})$ follows identically with $\mathcal{R}_{\mathrm{bb}}$ and $K_{\mathrm{b}}$ replacing $\mathcal{R}_{\mathrm{aa}}$ and $K_{\mathrm{a}}$.
	
	\emph{Determinant of $\mathbf{R}_{\mathrm{aba}'\mathrm{b}'}$}:
	Partitioning the full block matrix in \eqref{eq:R_aba'b'_app} as $\begin{bmatrix} \mathbf{R}_{\mathrm{ab}}^{(2)} & \mathbf{B}^{(2)} \\ \mathbf{C}^{(2)} & \mathbf{R}_{\mathrm{a}'\mathrm{b}'} \end{bmatrix}$ with $\mathbf{R}_{\mathrm{ab}}^{(2)} = {\begin{bmatrix} \mathcal{R}_{\mathrm{aa}} & \mathcal{R}_{\mathrm{ab}} \\ \mathcal{R}_{\mathrm{ba}} & \mathcal{R}_{\mathrm{bb}} \end{bmatrix}}$, $\mathbf{B}^{(2)} = {\begin{bmatrix} \mathcal{R}_{\mathrm{aa}'} & \mathcal{R}_{\mathrm{ab}'} \\ \mathcal{R}_{\mathrm{ba}'} & \mathcal{R}_{\mathrm{bb}'} \end{bmatrix}}$, and $\mathbf{C}^{(2)} = (\mathbf{B}^{(2)})^{\mathsf{H}}$, and applying the Schur complement with respect to the pivot block $\mathbf{R}_{\mathrm{a}'\mathrm{b}'}$ yields \eqref{eq:det_Raba'b'}.
	
	\subsection{{Covariance and Field-Response Derivatives}}\label{sec:position_derivatives}
	{
		For $u_n\in\{x_n,y_n\}$, differentiating~\eqref{eq:bar_R} gives
		\begin{align}
			\frac{\partial\bar{\mathbf R}}{\partial u_n}
			=\sigma_b^2\left(\frac{\partial\mathbf A}{\partial u_n}\mathbf A^{\mathsf H}
			+\mathbf A\left(\frac{\partial\mathbf A}{\partial u_n}\right)^{\mathsf H}\right). \label{eq:position_cov_derivative}
		\end{align}
		The derivative matrix $\partial\mathbf A/\partial u_n$ is zero outside row $n$, whose entries follow from~\eqref{eq:steering_2d} as
		\begin{align}
			\left[\frac{\partial\mathbf A}{\partial x_n}\right]_{n,p}
			&=j\frac{2\pi}{\lambda}\sin\theta_p\cos\phi_p[\mathbf A(\mathbf U)]_{n,p},\notag\\
			\left[\frac{\partial\mathbf A}{\partial y_n}\right]_{n,p}
			&=j\frac{2\pi}{\lambda}\cos\theta_p[\mathbf A(\mathbf U)]_{n,p}. \label{eq:position_field_derivative}
		\end{align}
	}
\end{appendices}

\bibliographystyle{IEEEtran}
\bibliography{IEEEabrv,Ref1}

@STRING{IEEE_J_VT         = "{IEEE} Trans. Veh. Technol."}

@STRING{IEEE_J_JSAC       = "{IEEE} J. Sel. Areas Commun."}

@STRING{IEEE_J_COM        = "{IEEE} Trans. Commun."}

@STRING{IEEE_J_WCOM       = "{IEEE} Trans. Wireless Commun."}

@STRING{IEEE_J_IFS        = "{IEEE} Trans. Inf. Forensics Security"}

@BOOK{CoverThomas2006,
  author={Cover, Thomas M. and Thomas, Joy A.},
  title={Elements of Information Theory},
  edition={2},
  publisher={Wiley},
  address={Hoboken, NJ, USA},
  year={2006},
  doi={10.1002/047174882X}
}

@ARTICLE{mimo25tw,
	author={Tang, Jie and Wen, Hong and Song, Huan-Huan and Jiao, Long and Zeng, Kai},
	journal={IEEE Internet Things J.}, 
	title={Sharing Secrets via Wireless Broadcasting: A New Efficient Physical Layer Group Secret Key Generation for Multiple IoT Devices}, 
	year={2022},
	volume={9},
	number={16},
	pages={15228-15239},
}

@ARTICLE{6GKAF2024,
  author={Kihero, Abuu B. and Furqan, Haji M. and Sahin, M. M. and Arslan, Hüseyin},
  journal={IEEE Wireless Commun.}, 
  title={{6G} and Beyond Wireless Channel Characteristics for Physical Layer Security: Opportunities and Challenges}, 
  year={2024},
  month={Jun.},
  volume={31},
  number={3},
  pages={295--301},
  doi={10.1109/MWC.002.2300002},}

@ARTICLE{PLS2023,
  author={Xie, Ning and Zhang, JiaHeng and Zhang, QiHong},
  journal={IEEE Network}, 
  title={Security Provided by the Physical Layer in Wireless Communications}, 
  year={2023},
  month={Sep.},
  volume={37},
  number={5},
  pages={42-48},
}

@ARTICLE{Maurer1993,
  author={Maurer, U.M.},
  journal={IEEE Trans. Inf. Theory}, 
  title={Secret key agreement by public discussion from common information}, 
  year={1993},
  month={May},
  volume={39},
  number={3},
  pages={733-742},
}

@ARTICLE{HLQH24,
  author={Hu, Lei and Li, Guyue and Qian, Xuewen and Hu, Aiqun and Ng, Derrick Wing Kwan},
  journal={IEEE Trans. Wireless Commun.}, 
  title={Reconfigurable Intelligent Surface-Assisted Secret Key Generation in Spatially Correlated Channels}, 
  year={2024},
  month={Mar.},
  volume={23},
  number={3},
  pages={2153-2166},
}

@ARTICLE{LMZW25,
  author={Li, Dongming and Ma, Wanting and Zhou, Fuhui and Wu, Qihui and Ng, Derrick Wing Kwan},
  journal=IEEE_J_IFS, 
  title={Physical-Layer Key Generation Efficient Beamspace Adaptations in {5G} New Radio}, 
  year={2025},
  month={Jan.},
  volume={20},
  number={},
  pages={1535--1550},
}

@ARTICLE{LHCH26,
	author={Liu, Kexin and Huang, Kaizhi and Chu, Zheng and Hu, Xiaoyan and Jin, Liang and Wang, Feihu and Pei, Xinglong},
	journal={IEEE Trans. Veh. Technol.}, 
	title={Secure Integrated Communications and Sensing Using Physical Layer Key Generation}, 
	year={2026},
	volume={75},
	number={8},
	pages={18513-18518},
}

@ARTICLE{LCZCDM25,
	author={Lu, Tianyu and Chen, Liquan and Zhang, Junqing and Chen, Chen and Duong, Trung Q. and Matthaiou, Michail},
	journal={IEEE T. Inf. Foren. Sec.}, 
	title={Precoding Design for Key Generation in Extremely Large-Scale MIMO Near-Field Multi-User Systems}, 
	year={2025},
	volume={20},
	number={},
	pages={10572-10587},
}

@ARTICLE{21JiTVT,  
author={Ji, Zijie and Yeoh, Phee Lep and Zhang, Deyou and Chen, Gaojie and Zhang, Yan and He, Zunwen and Yin, Hao and Li, Yonghui},  
journal=IEEE_J_VT,   
title={Secret Key Generation for Intelligent Reflecting Surface Assisted Wireless Communication Networks},
month={Jan.},   
year={2021},  
volume={70},  
number={1},  
pages={1030--1034},  
doi={10.1109/TVT.2020.3045728}}

@article{2021Sum,
  title={Sum Secret Key Rate Maximization for {TDD} Multi-User Massive {MIMO} Wireless Networks},
  author={Li, Guyue and Sun, Chen and Jorswieck, Eduard A. and Zhang, Junqing and Hu, Aiqun and Chen, You},
  journal={IEEE Trans. Inf. Forensics Security},
  volume={16},
  pages={968--982},
  year={2021},
  doi={10.1109/TIFS.2020.3026466},
}

@ARTICLE{li2018high,
  author={Li, Guyue and Hu, Aiqun and Zhang, Junqing and Peng, Linning and Sun, Chen and Cao, Daming},
  journal=IEEE_J_COM, 
  title={High-Agreement Uncorrelated Secret Key Generation Based on Principal Component Analysis Preprocessing}, 
  month={Jul.},
  year={2018},
  volume={66},
  number={7},
  pages={3022--3034},
  doi={10.1109/TCOMM.2018.2814607}}

@Article{ZhangReview,
  author  = {Zhang, Junqing and Duong, Trung Q. and Marshall, Alan and Woods, Roger},
  title   = {Key Generation From Wireless Channels: A Review},
  journal = {IEEE Access},
  year    = {2016},
  volume  = {4},
  pages   = {614--626},
  doi     = {10.1109/ACCESS.2016.2521718},
}

@ARTICLE{22Sum-RIS,  
	author={Li, Guyue and Sun, Chen and Xu, Wei and Renzo, Marco Di and Hu, Aiqun},  
	journal=IEEE_J_IFS,   
	title={On Maximizing the Sum Secret Key Rate for Reconfigurable Intelligent Surface-Assisted Multiuser Systems}, 
	month = {Dec.},  
	year={2022},  
	volume={17},  
	number={},  pages={211--225},  doi={10.1109/TIFS.2021.3138612}
}

@ARTICLE{LCHW17,
  author={Liu, Yiliang and Chen, Hsiao-Hwa and Wang, Liangmin},
  journal={IEEE Commun. Surv. Tut.}, 
  title={Physical Layer Security for Next Generation Wireless Networks: Theories, Technologies, and Challenges}, 
  year={2017},
  month={1st Quart.},
  volume={19},
  number={1},
  pages={347--376},
  doi={10.1109/COMST.2016.2598968},
}

@ARTICLE{GCXM21,
  author={Guo, Dengke and Cao, Kuo and Xiong, Jun and Ma, Dongtang and Zhao, Haitao},
  journal={IEEE Internet Things J.}, 
  title={A Lightweight Key Generation Scheme for the {Internet} of {Things}}, 
  year={2021},
  month={Aug.},
  volume={8},
  number={15},
  pages={12137-12149},}

@ARTICLE{ZZHX22,
  author={Zhao, Hong and Zhang, Yuexin and Huang, Xinyi and Xiang, Yang and Su, Chunhua},
  journal={IEEE Internet Things J.}, 
  title={A Physical-Layer Key Generation Approach Based on Received Signal Strength in Smart Homes}, 
  year={2022},
  month={Apr.},
  volume={9},
  number={7},
  pages={4917-4927},
}

@ARTICLE{EKB21,
  author={Ebrahimi, Najme and Kim, Hun-Seok and Blaauw, David},
  journal={IEEE Trans. Microw. Theory Techn.}, 
  title={Physical Layer Secret Key Generation Using Joint Interference and Phase Shift Keying Modulation}, 
  year={2021},
  month={May},
  volume={69},
  number={5},
  pages={2673--2685},
  doi={10.1109/TMTT.2021.3058183},
}

@ARTICLE{DTAE23,
  author={Diamant, Roee and Tomasin, Stefano and Ardizzon, Francesco and Eccher, Davide and Casari, Paolo},
  journal=IEEE_J_IFS, 
  title={Secret Key Generation From Route Propagation Delays for Underwater Acoustic Networks}, 
  year={2023},
  month={May},
  volume={18},
  number={},
  pages={3318--3333},
}

@ARTICLE{ARCI93,
  author={Ahlswede, R. and Csiszar, I.},
  journal={IEEE Trans. Inf. Theory}, 
  title={Common Randomness in Information Theory and Cryptography. {I}. Secret Sharing}, 
  year={1993},
  month={Jul.},
  volume={39},
  number={4},
  pages={1121--1132},
  doi={10.1109/18.243431},
}

@ARTICLE{JWWAT19,
  author={Jiao, Long and Wang, Ning and Wang, Pu and Alipour-Fanid, Amir and Tang, Jie and Zeng, Kai},
  journal={ IEEE Wirel. Commun.}, 
  title={Physical Layer Key Generation in {5G} Wireless Networks}, 
  year={2019},
  month={Oct.},
  volume={26},
  number={5},
  pages={48-54},
}

@ARTICLE{LXXJH21,
  author={Li, Guyue and Xu, Yinghao and Xu, Wei and Jorswieck, Eduard and Hu, Aiqun},
  journal=IEEE_J_IFS, 
  title={Robust Key Generation With Hardware Mismatch for Secure {MIMO} Communications}, 
  year={2021},
  month={Nov.},
  volume={16},
  number={},
  pages={5264--5278},
}

@ARTICLE{LCZCH23,
  author={Lu, Tianyu and Chen, Liquan and Zhang, Junqing and Chen, Chen and Hu, Aiqun},
  journal=IEEE_J_IFS, 
  title={Joint Precoding and Phase Shift Design in Reconfigurable Intelligent Surfaces-Assisted Secret Key Generation}, 
  year={2023},
  month={Apr.},
  volume={18},
  number={},
  pages={3251--3266},
}

@ARTICLE{ZhuMA23,
  author={Zhu, Lipeng and Ma, Wenyan and Zhang, Rui},
  journal={IEEE Commun. Mag.},
  title={Movable Antennas for Wireless Communication: Opportunities and Challenges},
  year={2024},
  volume={62},
  number={6},
  pages={114--120},
  month={Jun.},
}

@ARTICLE{Matsumine25,
  author={Matsumine, Toshiki and Ochiai, Hideki and Shikata, Junji},
  journal={IEEE Wireless Commun. Lett.},
  title={Enhancing Physical-Layer Secret Key Generation Through Random Antenna Positioning},
  year={2026},
  volume={15},
  pages={3149--3153},
  doi={10.1109/LWC.2026.3693289},
}

@ARTICLE{XuMA,
  author={Xu, Bokai and Zhang, Jiayi and Liu, Ziheng and Wang, Zhe and Ai, Bo and Ng, Derrick Wing Kwan},
  journal={IEEE Trans. Wireless Commun.},
  title={Asynchronous Distributed Beamforming for Beyond-Diagonal {RIS}-Aided Movable Antenna Systems},
  year={2026},
  volume={25},
  pages={4737--4751},
  doi={10.1109/TWC.2025.3613556},
}

@ARTICLE{MATutorial25,
  author={Zhu, Lipeng and Ma, Wenyan and Mei, Weidong and Zeng, Yong and Wu, Qingqing and Ning, Boyu and Xiao, Zhenyu and Shao, Xiaodan and Zhang, Jun and Zhang, Rui},
  journal={IEEE Commun. Surveys Tuts.},
  title={A Tutorial on Movable Antennas for Wireless Networks},
  year={2026},
  month={Fourth Quarter},
  volume={28},
  number={4},
  pages={3002--3054},
  doi={10.1109/COMST.2025.3546373},
}

@ARTICLE{MZSZ26,
	author={Ma, Wenyan and Zhu, Lipeng and Shao, Xiaodan and Zhang, Rui},
	journal={IEEE Trans. Wireless Commun.}, 
	title={{3-D} Trajectory Optimization for Robust Direction Sensing in Movable Antenna Systems}, 
	year={2026},
	volume={25},
	number={},
	pages={19677-19691},
}

@ARTICLE{ModelingMA23,
  author={Zhu, Lipeng and Ma, Wenyan and Zhang, Rui},
  journal={IEEE Trans. Wireless Commun.},
  title={Modeling and Performance Analysis for Movable Antenna Enabled Wireless Communications},
  year={2024},
  month={Jun.},
  volume={23},
  number={6},
  pages={6234--6250},
  doi={10.1109/TWC.2023.3330887},
}

@ARTICLE{6DQDLSZ26,
	author={Qian, Yanzhi and Ding, Jingze and Liu, Sujie and Shao, Xiaodan and Zheng, Beixiong and Zhu, Lipeng},
	journal={IEEE Communications Letters}, 
	title={{6DMA}-Assisted Secure Wireless Communications}, 
	year={2026},
	volume={30},
	number={},
	pages={1499-1503},
}

@ARTICLE{MaMIMO24,
  author={Ma, Wenyan and Zhu, Lipeng and Zhang, Rui},
  journal={IEEE Trans. Wireless Commun.},
  title={{MIMO} Capacity Characterization for Movable Antenna Systems},
  year={2024},
  volume={23},
  number={4},
  pages={3392-3407},
}

@ARTICLE{YZZ26,
	author={Yan, Ge and Zhu, Lipeng and Zhang, Rui},
	journal={IEEE Trans. Commun.}, 
	title={Movable Antenna Aided Multiuser Communications: Antenna Position Optimization Based on Statistical Channel Information}, 
	year={2026},
	volume={74},
	number={},
	pages={1793-1810},
}

@ARTICLE{xlgyxt26,
	author={Xie, Hao and Li, Dong and Gu, Bowen and Yu, Xianhua and Xu, Yongjun and Tellambura, Chintha},
	journal={IEEE Trans. Commun.}, 
	title={Movable Antenna-Aided Wireless Systems: Concurrent or Cumulative Movement?}, 
	year={2026},
	volume={74},
	number={},
	pages={8829-8842},
}

@INPROCEEDINGS{ChengSecure24,
  author={Cheng, Zhenqiao and Li, Nanxi and Zhu, Jianchi and She, Xiaoming and Ouyang, Chongjun and Chen, Peng},
  booktitle={Proc. IEEE Int. Conf. Acoustics, Speech and Signal Process. (ICASSP)},
  title={Enabling Secure Wireless Communications via Movable Antennas},
  year={2024},
  pages={9186-9190},
}

@ARTICLE{HuSPL24,
  author={Hu, Guojie and Wu, Qingqing and Xu, Kui and Si, Jiangbo and Al-Dhahir, Naofal},
  journal={IEEE Signal Process. Lett.},
  title={Secure Wireless Communication via Movable-Antenna Array},
  year={2024},
  volume={31},
  pages={516-520},
}

@ARTICLE{TangMIMO25,
  author={Tang, Jun and Pan, Cunhua and Zhang, Yang and Ren, Hong and Wang, Kezhi},
  journal={IEEE Trans. Commun.},
  title={Secure MIMO Communication Relying on Movable Antennas},
  year={2025},
  volume={73},
  number={4},
  pages={2159-2175},
}

@INPROCEEDINGS{ChengNoCSI25,
  author={Cheng, Zhenqiao and Ouyang, Chongjun and Zhang, Xingqi},
  booktitle={Proc. IEEE Int. Conf. Acoustics, Speech and Signal Process. (ICASSP)},
  title={Movable Antenna Aided Physical Layer Security with No Eavesdropper CSI},
  year={2025},
}

@ARTICLE{HuTMC24,
  author={Hu, Guojie and Wu, Qingqing and Xu, Donghui and Xu, Kui and Si, Jiangbo and Cai, Yunlong and Al-Dhahir, Naofal},
  journal={IEEE Trans. Mobile Comput.},
  title={Movable Antennas-Assisted Secure Transmission Without Eavesdroppers' Instantaneous CSI},
  year={2024},
  volume={23},
  number={12},
  pages={14263-14279},
}

@ARTICLE{FengPLS25,
  author={Feng, Zhiyong and Zhao, Yujia and Yu, Kan and Li, Dong},
  journal={IEEE Trans. Commun.},
  title={Movable Antenna Empowered Physical Layer Security With Eavesdropper Location Uncertainty: Joint Optimization of Beamforming and Antenna Positions},
  year={2025},
  volume={73},
  number={12},
  pages={13708-13724},
}

@ARTICLE{LiuCovert25,
  author={Liu, Pengpeng and Si, Jiangbo and Cheng, Zihao and Li, Zan and Hu, Hang},
  journal={IEEE Wireless Commun. Lett.},
  title={Movable-Antenna Enabled Covert Communication},
  year={2025},
  volume={14},
  number={2},
  pages={280-284},
}

@ARTICLE{MaoCovert25,
  author={Mao, Haobin and Pi, Xiangyu and Zhu, Lipeng and Xiao, Zhenyu and Xia, Xiang-Gen and Zhang, Rui},
  journal={IEEE Wireless Commun. Lett.},
  title={Sum Rate Maximization for Movable Antenna Enhanced Multiuser Covert Communications},
  year={2025},
  volume={14},
  number={3},
  pages={611-615},
}

@ARTICLE{LeISAC25,
  author={Le, Hoang Hung and Nguyen, Hoang Huy and Nguyen, Cong Luong and Pham, Quoc-Viet and Niyato, Dusit and Nguyen, Tien Hoa},
  journal={IEEE Trans. Veh. Technol.},
  title={Beamforming Design for Physical Security in Movable Antenna-Aided ISAC Systems: A Reinforcement Learning Approach},
  year={2025},
  volume={74},
  number={11},
  pages={18163-18167},
}

@ARTICLE{BSPL26,
	author={Biswas, Sudip and Singh, Keshav and Patel, Ravi and Pan, Cunhua and Li, Chih-Peng},
	journal={IEEE Trans. Wireless Commun.}, 
	title={Joint Beamforming, RIS Configuration, and Antenna Positioning for Active RIS-Assisted ISAC With Movable-Antenna Arrays}, 
	year={2026},
	volume={25},
	number={},
	pages={14517-14534},
}

@ARTICLE{CZGWNZWZ26,
	author={Chen, Guangyi and Zhang, Ruoyu and Guan, Xinrong and Wu, Qingqing and Ning, Boyu and Zhang, Yu and Wu, Wen and Zhang, Rui},
	journal={IEEE Trans. Wireless Commun.}, 
	title={Movable Antenna-Enabled MIMO Integrated Sensing and Communication: A Unified Mutual Information Framework}, 
	year={2026},
	volume={25},
	number={},
	pages={14440-14454},
}

@ARTICLE{BZCLYJS26,
	author={Bian, Hongyi and Zhao, Feilong and Chen, Minghao and Liu, Lingling and Yao, Yu and Jiang, Hao and Shu, Feng},
	journal={IEEE Transactions on Network Science and Engineering}, 
	title={Movable Antenna-Enabled Secure Transmission for Active {RIS}-Aided {ISAC} Systems}, 
	year={2026},
	volume={13},
	number={},
	pages={6325-6344},
}

@ARTICLE{LiPilot25,
  author={Li, Danqi and Yan, Shihao and Zhang, Jia and Sun, Jiande and Shu, Feng},
  journal={IEEE Wireless Commun. Lett.},
  title={Optimal Pilot Length for Key Generation in Correlated Wireless Channels},
  year={2024},
  volume={13},
  number={10},
  pages={2767-2771},
  doi={10.1109/LWC.2024.3444821},
}

@INPROCEEDINGS{ZhangSlowFading,
  author={Zhang, Yuang and Chen, Liquan and Song, Yufan and Lv, Yan},
  booktitle={Proc. IEEE Wireless Commun. Networking Conf. (WCNC)},
  title={Secret Key Capacity of Key Generation in the Slow-Fading Wireless Channels},
  address={Milan, Italy},
  year={2025},
  pages={1-6},
}

@ARTICLE{Zhangjun_cor,
	author={Zhang, Junqing and He, Biao and Duong, Trung Q. and Woods, Roger},
	journal={IEEE Commun. Lett.}, 
	title={On the Key Generation From Correlated Wireless Channels}, 
	year={2017},
	volume={21},
	number={4},
	pages={961-964},
}

@ARTICLE{HuangPLKG,
	author={Huang, Zhiyu and Li, Guyue and Xu, Hao and Ng, Derrick Wing Kwan},
	journal=IEEE_J_JSAC, 
	title={Fluid Antenna System-Assisted Physical Layer Secret Key Generation}, 
	year={2026},
	volume={44},
	number={},
	pages={4429--4443},
	doi={10.1109/JSAC.2026.3679311},
}

@ARTICLE{PlaneWave26,
	author={Lu, Haiquan and Zeng, Yong and Ma, Shaodan and Li, Bin and Jin, Shi and Zhang, Rui},
	journal=IEEE_J_WCOM, 
	title={Wireless Communication for Low-Altitude Economy With UAV Swarm Enabled Two-Level Movable Antenna System}, 
	year={2026},
	volume={25},
	number={},
	pages={16463--16479},
}

@ARTICLE{MYNMCF21,
	author={Matthaiou, Michail and Yurduseven, Okan and Ngo, Hien Quoc and Morales-Jimenez, David and Cotton, Simon L. and Fusco, Vincent F.},
	journal={IEEE Commun. Mag.}, 
	title={The Road to 6G: Ten Physical Layer Challenges for Communications Engineers}, 
	year={2021},
	volume={59},
	number={1},
	pages={64-69},
}

%
%
	%

\end{document}